%% file: main.tex
\documentclass[letterpaper]{article} 
\usepackage{aaai24}  
\usepackage{times}  
\usepackage{helvet}  
\usepackage{courier}  
\usepackage[hyphens]{url}  
\usepackage{graphicx} 
\usepackage{natbib}  
\usepackage{caption} 
\usepackage{algorithm}
\usepackage{algorithmicx}
\usepackage[noend]{algpseudocode}
\algrenewcommand\algorithmiccomment[1]{\hfill\textcolor{blue}{\commentsymbol{} #1}}

\usepackage{newfloat}
\usepackage{listings}
\DeclareCaptionStyle{ruled}{labelfont=normalfont,labelsep=colon,strut=off} 
\floatstyle{ruled}
\newfloat{listing}{tb}{lst}{}
\floatname{listing}{Listing}

\usepackage{import}
\usepackage{amssymb}
\usepackage{amsmath}
\usepackage{booktabs}

\usepackage{amsthm}
\newtheorem{theorem}{Theorem}
\theoremstyle{definition}
\newtheorem{example}{Example}

\usepackage{tikz,pgffor}
\usepackage{ifthen}
\usetikzlibrary{arrows,backgrounds,calc}
\usetikzlibrary{automata,positioning,arrows,shapes,math,arrows.meta,decorations.pathmorphing}
\tikzstyle{min}=[thick,circle,draw,minimum size=1.4em,inner sep=0em,text centered]
\tikzstyle{dec}=[circle,draw,fill,minimum size=.8ex,inner sep=0em]

\input{macros}

\graphicspath{ {./plots/} }

\title{Optimizing Local Satisfaction of Long-Run Average Objectives\\ in Markov Decision Processes}
\author {
David Kla\v{s}ka\textsuperscript{\rm 1},
Anton\'{\i}n Ku\v{c}era\textsuperscript{\rm 1},
Vojt\v{e}ch K\r{u}r\textsuperscript{\rm 1}, 
V\'{\i}t Musil\textsuperscript{\rm 1},
Vojt\v{e}ch \v{R}eh\'{a}k\textsuperscript{\rm 1} 
}
\affiliations {
    \textsuperscript{\rm 1}Masaryk University, Brno, Czechia\\
    david.klaska@mail.muni.cz, tony@fi.muni.cz, vojtech.kur@mail.muni.cz, musil@fi.muni.cz, rehak@fi.muni.cz
}

\begin{document}

\maketitle

\begin{abstract}
Long-run average optimization problems for Markov decision processes (MDPs) require constructing policies with optimal steady-state behavior, i.e., optimal limit frequency of visits to the states.
However, such policies may suffer from \emph{local instability}, i.e., the frequency of states visited in a bounded time horizon along a run differs significantly from the limit frequency. In this work, we propose an efficient algorithmic solution to this problem.
\end{abstract}

\import{sections}{intro}

\import{sections}{model}

\import{sections}{evaluate}
\import{sections}{optimize}

\import{sections}{experiments}

\newpage
\section*{Acknowledgments}

Research was sponsored by the Army Research Office and accomplished under
Grant Number W911NF-21-1-0189.

\noindent
\textit{Disclaimer.}\quad The views and conclusions contained in this document are those of the authors and should not be interpreted as representing the official policies, either
expressed or implied, of the Army Research Office or the U.S.\ Government. The
U.S.\ Government is authorized to reproduce and distribute reprints for
Government purposes notwithstanding any copyright notation herein.

Vojt\v{e}ch K\r{u}r received funding from the European Union's Horizon Europe program under the Grant Agreement No.\ 101087529.  V\'{\i}t Musil was supported by the Czech Science Foundation grant GA23-06963S.

\bibstyle{aaai24}
\bibliography{str-long,concur}

\newpage
\begin{center}
    \huge\bf Appendix
\end{center}
\appendix

\import{suppsections}{hardness}

\import{suppsections}{algorithms}
\import{suppsections}{experiments}

\end{document}

%% file: macros.tex
\usepackage{tikz,pgffor}
\usetikzlibrary{arrows,backgrounds,calc}
\usetikzlibrary{automata,positioning,arrows,shapes,math,arrows.meta,decorations.pathmorphing}
\tikzstyle{min}=[thick,circle,draw,minimum size=1.3em,inner sep=0em,text centered]

\newcommand{\suppl}{Appendix}

\newcommand{\Nset}{\mathbb{N}}

\newcommand{\Rset}{\mathbb{R}}
\newcommand{\Exp}{\mathbb{E}}
\newcommand{\Var}{\mathrm{Var}}
\newcommand{\Inv}{\mathbb{I}}

\newcommand{\SD}{\textit{SD}}

\newcommand{\Mem}{M}
\newcommand{\prob}{\mathbb{P}}
\newcommand{\Prob}{\textit{Prob}}

\newcommand{\Init}{\mathit{Init}}
\newcommand{\Freq}{\mathit{Freq}}
\newcommand{\Penalty}{\mathit{Penalty}}
\newcommand{\Satisfy}{\mathit{Satisfy}}

\newcommand{\Obj}{\textit{Obj}}

\newcommand{\RT}{\mathit{RT}}

\newcommand{\LBad}{\textit{L-Badness}}

\newcommand{\Dist}{\mathit{Dist}}
\newcommand{\Distance}{\mathit{Distance}}

\newcommand{\Comb}{\mathit{Comb}}

\newcommand{\ag}[1]{\overline{#1}}

\newcommand{\norm}[1]{|\!|#1|\!|}

\newcommand{\calL}{\mathcal{L}}

\newcommand{\PSPACE}{$\mathsf{PSPACE}$}
\newcommand{\NP}{$\mathsf{NP}$}
\newcommand{\LE}{$\mathsf{LocalEval}$}
\newcommand{\LS}{$\mathsf{LocalSynt}$}

\usepackage{xspace}
\makeatletter
\DeclareRobustCommand\onedot{\futurelet\@let@token\@onedot}
\def\@onedot{\ifx\@let@token.\else.\null\fi\xspace}

\def\ie{i.e\onedot}

\makeatother

%% file: sections/intro.tex
\section{Introduction}
\label{sec-intro}


A \emph{long-run average objective} for a Markov decision process (MDP) $D$ is a property depending on the proportion of time (frequency) spent in the individual states of~$D$. Typical examples of such properties include
\begin{itemize}
   \item the total frequency of visits to ``bad'' states is  ${\leq} 0.05$;
   \item the state frequency vector is equal to a given vector $\nu$.
\end{itemize}
The existing works on long-run average optimization (see Related Work) concentrate on constructing a strategy $\sigma$ such that the Markov chain $D^\sigma$ obtained by applying $\sigma$ to $D$ is irreducible and the \emph{invariant} (also called \emph{steady-state}) distribution $\Inv_\sigma$ achieves the objective\footnote{Recall that by ergodic theorem \cite{Norris:book}, the invariant distribution is the vector of limit state frequencies computed for longer and longer prefixes of runs.}. Unfortunately, the existing algorithms cannot influence the \emph{local stability} of the invariant distribution along a run.

More concretely, for a given time horizon~$n$, consider the \emph{local frequency} $\Freq_n$ of states sampled from $n$ consecutive states along a run, starting at a randomly chosen \emph{pivot position}  (we refer to Section~\ref{sec-model} for precise definitions). The local stability of the invariant distribution is the probability that $\Freq_n$ stays ``close'' to  $\Inv_\sigma$. If the local stability is low, then the probability of achieving the considered objective \emph{locally} (i.e., within the prescribed time horizon) is also low, and this may lead to severe problems in many application scenarios. 

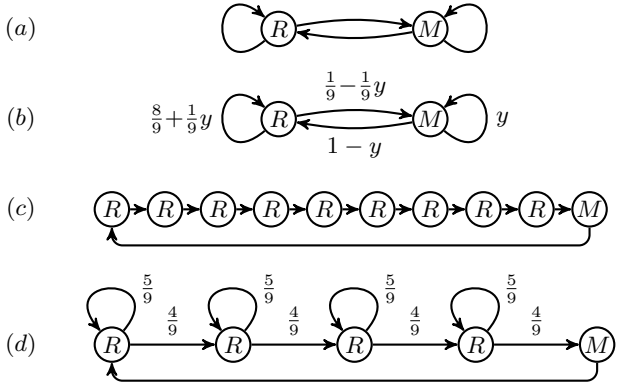
\begin{figure}[t]\centering
\begin{tikzpicture}[x=4.2cm, y=2.5cm, >=stealth', scale=0.48,font=\footnotesize]
    \foreach \y in {0,-1}{%
       \node[min] (R\y) at (0,\y) {$R$};
       \node[min] (M\y) at (1,\y) {$M$};
       \draw[->,thick] (R\y) to [in=170,out=10] node[above] {\ifthenelse{\y=0}{}{$\frac{1}{9}{-}\frac{1}{9}y$}} (M\y);
       \draw[->,thick] (M\y) to [in=350,out=190] node[below] {\ifthenelse{\y=0}{}{$1-y$}} (R\y);
       \path (M\y) edge [out=-40,in=40,thick, looseness=8, ->] node[right] {\ifthenelse{\y=0}{}{$y$}} (M\y);
       \path (R\y) edge [out=-140,in=140,thick, looseness=8, ->] node[left] {\ifthenelse{\y=0}{}{$\frac{8}{9}{+}\frac{1}{9}y$}} (R\y);
       \node (l\y) at (-1.7,\y) {\ifthenelse{\y=0}{$(a)$}{$(b)$}};
    }
    \coordinate (a) at (-1.1,-2); 
    \foreach \x/\y in {0/0, 1/0.35, 2/.7, 3/1.05, 4/1.4, 5/1.75, 6/2.1, 7/2.45, 8/2.8, 9/3.15}{%
       \node[min] (N\x) at ($(a) +(\y,0)$) {\ifthenelse{\x=9}{$M$}{$R$}};
    }
    \foreach \x/\y in {0/1, 1/2, 2/3, 3/4, 4/5, 5/6, 6/7, 7/8, 8/9}{%
       \draw[->,thick] (N\x) to (N\y);
    }
    \draw[->,thick, rounded corners] (N9) -- ($(N9) +(0,-.4)$) -| (N0);
    \node (l) at (-1.7,-2) {$(c)$};
    \coordinate (a) at (-1.1,-3.5); 
    \foreach \x/\y in {0/0, 1/0.8, 2/1.6, 3/2.4, 4/3.2}{%
       \node[min] (N\x) at ($(a) +(\y,0)$) {\ifthenelse{\x=4}{$M$}{$R$}};
    }
    \foreach \x/\y in {0/1, 1/2, 2/3, 3/4}{%
       \draw[->,thick] (N\x) to node[above] {$\frac{4}{9}$} (N\y);
       \path (N\x) edge [out=50,in=130,thick, looseness=8, ->] node[right=1.5ex] {$\frac{5}{9}$} (N\x);
    }
    \draw[->,thick, rounded corners] (N4) -- ($(N4) +(0,-.4)$) -| (N0);
    \node (l) at (-1.7,-3.5) {$(d)$};
\end{tikzpicture}
\caption{For the graph~(a), the memoryless strategy $\sigma_{y}$ of~(b) achieves $\Inv_{\sigma_{y}} = \nu = (0.9,0.1)$ for all $y \in [0,1)$, but $\Prob^{\sigma_{y}}[\Freq_{10} {=} \nu] \leq 0.43$ for all  $y \in [0,1)$. The deterministic finite-memory strategy $\pi$ of~(c) achieves $\Inv_{\pi} = \nu$ and $\Prob^{\pi}[\Freq_{10} {=} \nu] = 1$ at the cost of large memory. The randomized finite-memory strategy $\eta$ of~(d) achieves $\Inv_{\eta} = \nu$ and $\Prob^{\eta}[\Freq_{10} {=} \nu] \approx 0.74$ with less memory.}
\label{fig-RM-system}
\end{figure}

\begin{example}
   \label{exa-RM-system}
Consider a system of Fig.~\ref{fig-RM-system}(a) that can be either in the running (R) or maintenance (M) state. A long-run sustainability of the system requires that the system is running for $90\%$ of time and the remaining $10\%$ is spent on maintenance. Hence, we aim at constructing a strategy $\sigma$ such that $\Inv_{\sigma} = \nu$, where $\nu(R) = 0.9$ and $\nu(M) = 0.1$. Ideally, the maintenance should be performed \emph{regularly}, i.e., the state $M$ should be visited once in $10$ consecutive states. That is, $\Freq_{10}$ should be equal to $\nu$ with high probability.

For every $y \in [0, 1)$, the memoryless strategy $\sigma_{y}$ of Fig.~\ref{fig-RM-system}(b) satisfies $\Inv_{\sigma_{y}} = \nu$. However, the probability of $\Prob^{\sigma_{y}}[\Freq_{10} {=} \nu]$ approaches \emph{zero} as $y \to 1$. The best result is achieved for $y = 0$, where this probability is ${\approx} 0.43$. Hence, even the best memoryless strategy may considerably degrade the reliability of the system.

The simple deterministic strategy $\pi$ of Fig.~\ref{fig-RM-system}(c) satisfies $\Inv_{\pi} = \nu$
and $\Prob^{\pi}[\Freq_{10} {=} \nu] = 1$. Note that $\pi$ needs $9$ memory states to ``count'' the repeated visits to $R$ before visiting~$M$. A ``tradeoff'' between memory size and the local satisfaction of the sustainability objective is achieved by the strategy $\eta$ of Fig.~\ref{fig-RM-system}(d) where $\Inv_{\eta} = \nu$ and  $\Prob^{\eta}[\Freq_{10} {=} \nu] \approx 0.74$.
\qed
\end{example}

Other examples of long-run average objectives where the local satisfaction/stability requirements rise naturally are 
\begin{itemize}
   \item \emph{critical supply delivery} (see, e.g., \cite{Skwirzynski-multiuser-comm,Lazar:optimal-flow-queues}), where a bundle of items with limited lifespan should be delivered with a given frequency $f$. A high level of local instability of the frequency causes a high probability of early/late deliveries that are both undesirable (early deliveries lead to wasting the items that are not consumed before expiration, and late deliveries lead to a shortage of items).    
   \item \emph{dependability}, i.e., an upper bound on failure frequency (see, e.g., \cite{BL:MDP-failure-rate, BLF:non-ergodic-MDP-failure-rate}). If this bound is locally violated with considerable probability, a user may interpret this as a violation of the dependability guarantee. For example, consider a device supposed to fail at most once in a month \emph{on average} during the device lifetime. If the device fails twice in two weeks with probability $0.2$ (which is possible \emph{without} violating the guarantee on the long-run average failure frequency), the device is likely to be perceived as \emph{unreliable}. 
\end{itemize}
The above list of examples is not exhaustive. Scenarios documenting the importance of local satisfaction/stability can be found in every application area involving long-run average objectives.

\paragraph{Our Contribution}
Example~\ref{exa-RM-system} shows that optimizing the local satisfaction of long-run average objectives is non-trivial even for small \emph{graphs} (i.e., MDPs with no probabilistic choice) and optimal strategies may require memory of considerable size.  In this work, we formalize the notion of local satisfaction, examine its computational hardness, and design an efficient strategy synthesis algorithm for maximizing the local satisfaction of a given objective in a given MDP. The algorithm is evaluated on examples of non-trivial size. To the best of our knowledge, this is the first systematic study of the local stability of invariant distributions along runs in MDPs and the associated algorithmic problems. More concretely, our results can be summarized as follows:

\textbf{I.} We introduce an abstract class of long-run average objectives and precisely formulate the local optimization problem for a given objective and MDPs. We show that computing an optimal strategy is \NP-hard even for \emph{graphs}.

\textbf{II.}  We design a dynamic algorithm \LE\ for evaluating the local satisfaction of a given objective $\Obj$ achieved by a given finite-memory strategy $\sigma$. We show that, on the one hand, \LE\ substantially outperforms a naive algorithm based on depth-first search, but, on the other hand, \LE\ is not sufficiently efficient for purposes of automatic differentiation and gradient descent.

\textbf{III.}  We propose an efficient algorithm \LS\ for synthesizing a finite-memory strategy $\sigma$ maximizing the local satisfaction of a given $\Obj$ in a given MDP. 
   \LS\ is based on isolating three crucial features of $\sigma$ that influence the local satisfaction of $\Obj$: 
   \begin{description}
      \item[F1.] The ``appropriateness'' of $\Inv_\sigma$ for satisfying $\Obj$.
      \item[F2.] The ``regularity'' of $\sigma$, i.e., the stochastic stability of renewal times for certain families of states.
      \item[F3.] The ``level of determinism'' of $\sigma$. 
   \end{description}
   Subsequently, we design highly efficient evaluation functions for F1--F3 and optimize them jointly by gradient descent. We experimentally confirm the scalability of \LS\ and the expected impact of different F1--F3 prioritization on the properties of the constructed strategies.

\paragraph{Related Work}

The \emph{steady-state strategy synthesis problem}, i.e., the task of constructing a strategy for a given MDP achieving a given invariant distribution, has been solved in \cite{BBCFK:MDP-two-views}
(see also \cite{BBCFK:MDP-two-views-LMCS}) even for a more general class of multiple mean-payoff objectives. The constructed strategies may require infinite memory in general and can be computed in polynomial time. The problem of constructing a \emph{memoryless randomized} strategy achieving a given steady-state distribution has been considered in \cite{ABHH:steady-state-ergodic-MDP} for a subclass of \emph{ergodic} MDPs and in \cite{Velasquez-steady-state,ABAV:steady-state-synthesis-multichain} for general MDPs. A polynomial-time strategy synthesis algorithm based on linear programming is given in both cases. The problem of computing a \emph{deterministic} strategy achieving a given invariant distribution has been shown \NP-hard and solvable by integer programming in \cite{VASWA:deterministic-steady-state-JAR}. More recently, steady-state strategy synthesis under LTL constraints has been solved in \cite{Kretinsky:steady-state-LTL}.

Optimizing \emph{expected window mean-payoff} for MDP \cite{BGR:window-mean-payoff} is perhaps most related to the problem studied in this paper. Here, each MDP state is assigned a payoff collected when visiting the state. The task is to ensure that the average reward per visited state (mean-payoff) in a window of length $\ell$ sliding along a run reaches a given threshold within the window length. This can be seen as enforcing a form of ``local stability'' of the mean payoff along a run. The problem
is solvable in time polynomial in the size of MDP and $\ell$, and the algorithm relies on previous results achieved for 2-player games \cite{CHDRR:window-mean-payoff-games-IC}. This technique is not applicable in our setting (recall that the studied problem is \NP-hard even for graphs). 

In a broader perspective, 
there are also works studying the trade-offs between the overall expected performance (mean payoff) and some forms of stability measured by variances of appropriate random variables \cite{BCFK:performance-stability-JCSS}.

%% file: sections/model.tex
\section{The Model}
\label{sec-model}

We assume familiarity with basic notions of probability theory (probability distribution, expected value, conditional variance, etc.) and Markov chain theory. The set of all probability distributions over a finite set~$A$ is denoted by $\Dist(A)$. 

\paragraph{Markov chains}
A \emph{Markov chain} is a triple $C=(S,\Prob, \mu)$ where $S$ is  a finite set of states, 
$\Prob\colon S\times S \to [0,1]$ is a stochastic matrix  such that $\sum_{s' \in S}\Prob(s,s') = 1$ for every $s \in S$, and $\mu \in \Dist(S)$ is an initial distribution. 

A \emph{run} of $C$ is an infinite sequence $w = s_0,s_1,\ldots$ of states. We use $\prob_{\mu}$ to denote the probability measure in the standard probability space over the runs of~$C$ determined by $\Prob$ and $\mu$, and we use $\Init(w)$ to denote the initial state of $w$ (\ie, $\Init(w) = s_0$).

Let $s,t \in S$. We say that $t$ is \emph{reachable} from $s$ if the probability of visiting $t$ from $s$ is positive, i.e., \mbox{$\Prob^n(s,t) > 0$} for some $n\geq 0$ (recall that $\Prob^0$ is the identity matrix).



\paragraph{Markov decision processes (MDPs)} A \emph{Markov decision process (MDP)}\footnote{Our definition of MDPs is standard in the area of graph games. It is equivalent to the ``classical'' MDP definition where  \emph{actions} are used instead of stochastic vertices (see, e.g., \cite{Puterman:book}). For our purposes, the adopted definition is more convenient and leads to substantially simpler notation.} is a triple 
$D {=} (V,E,p)$ where $V$ is a finite set of \emph{vertices} partitioned into subsets $(V_N,V_S)$ of \emph{non-deterministic} and \emph{stochastic} vertices, $E \subseteq V {\times} V$ is a set of \emph{edges} s.t.{} every vertex has at least one out-going edge, and $p\colon V_S {\to} \Dist(V)$ is a \emph{probability assignment} s.t.{} $p(v)(v') {>} 0$ only if $(v,v') \in E$. We say $D$ is a \emph{graph} if $V_S {=} \emptyset$.

Outgoing edges in non-deterministic states are selected by a \emph{strategy}. The most general type of strategy is a \emph{history-dependent randomized (HR)} strategy where the selection is randomized and depends on the whole computational history. Since HR strategies require infinite memory, they are not apt for algorithmic purposes. Therefore, we restrict ourselves to a subclass of \emph{finite-memory randomized (FR)} strategies introduced in the next paragraph.

\paragraph{FR strategies}
Let $D = (V,E,p)$ be an MDP and $\Mem \neq \emptyset$ a finite set of \emph{memory states}. Intuitively, memory states are used to ``remember'' some information about the sequence of previously visited vertices. For a given pair $(v,m)$ where $v$ is a currently visited vertex and $m$ a current memory state, a strategy randomly selects a new pair $(v',m')$ such that $(v,v') \in E$. In general, the new memory state $m'$ may \emph{not} be uniquely determined by the chosen $v'$. If $v$ is stochastic, then $v'$ is selected with probability $p(v)(v')$, and the strategy randomly selects the new memory state $m'$. 

Formally, let $\alpha\colon V \to 2^{\Mem}$ be a \emph{memory allocation} assigning to every vertex $v$ a non-empty subset of memory states available in $V$. Let 
$\ag{V} = \{(v,m) \mid v \in V, m \in \alpha(v)\}$ be the set of \emph{augmented vertices}. A \emph{finite-memory (FR) strategy} is a function
$\sigma\colon \ag{V} \to \Dist(\ag{V})$ such that for all $(v,m) \in \ag{V}$ where $v \in V_S$ and every $(v,v')\in E$ we have that 
\[
\sum_{m' \in \alpha(v')}\sigma(v,m)(v',m') \ = \ p(v)(v')\,.
\]
An FR strategy is \emph{memoryless} (or \emph{Markovian}) if $\Mem$ is a singleton.  In the following,
we use $\ag{v}$ to denote an augmented vertex of the form $(v,m)$ for some $m\in\alpha(v)$. 

Every FR strategy $\sigma$ together with a probability distribution $\mu \in \Dist(\ag{V})$ determine the Markov chain $D^{\sigma} = (\ag{V},\Prob,\mu)$ where $\Prob(\ag{v},\ag{u}) = \sigma(\ag{v})(\ag{u})$. 

\paragraph{Invariant distributions}
Let $C=(S,\Prob, \mu)$ be a Markov chain. A \emph{bottom strongly connected component (BSCC)} of $C$ is a maximal $B \subseteq S$ such that $B$ is strongly connected and closed under reachable states, i.e., for all $s,t \in B$ and $r \in S$ we have that $t$ is reachable from $s$, and if $r$ is reachable from $s$, then $r \in B$. 

Let $B$ be a BSCC of $C$. For every $\nu \in \Dist(B)$, let $B^\nu$ be the Markov chain $(B,\Prob_B,\nu)$ where $\Prob_B$ is the restriction of $\Prob$ to  $B {\times} B$. Furthermore, let $\Inv_B \in \Dist(B)$ be the unique \emph{invariant distribution} satisfying $\Inv_B = \Inv_B \cdot \Prob_B$ (note that $\Inv_B$ is independent of $\nu$). By ergodic theorem \cite{Norris:book}, $\Inv_B$ is the limit frequency of visits to the states of $B$ along a run in~$B^\nu$. More precisely, let $w = s_0,s_1,\ldots$ be a run of $B^\nu$. For every $n \geq 1$, let $\Freq_n(w)\colon B \to [0,1]$ be the state frequency vector computed for the prefix of $w$ of length $n$, i.e., for every $s \in B$,
\[
    \Freq_n(w)(s) =  \#_s(s_0,\ldots,s_{n-1})/n    
\]
where $\#_s(s_0,\ldots,s_{n-1})$ is the number of occurrences of $s$ in $s_0,\ldots,s_{n-1}$. 
Let $\Freq(w) = \lim_{n\to \infty} \Freq_n(w)$. If the limit does not exist, we put $\Freq(w) = \vec{0}$. The ergodic theorem says  that $\prob^{\nu}[\Freq {=} \Inv_B] = 1$.

\paragraph{Long-run average objectives}

Let $D = (V,E,p)$ be an MDP. A \emph{long-run average objective} for $D$ is a function $\Obj\colon \Dist(V) \to \Rset^{\geq 0}$. Intuitively, for a given frequency of visits to $V$, the value of $\Obj$ specifies the ``badness'' of the frequency, i.e., a higher value of $\Obj(\mu)$ indicates that $\mu$ is ``less appropriate'' for achieving the objective encoded by $\Obj$. Two representative examples are given below.
\begin{itemize}
    \item For a given $\nu \in \Dist(V)$, let \mbox{$\Distance_\nu(\mu) = \norm{\mu - \nu}$}, where $\norm{\cdot}$ is a vector norm (such as $L_1$ or $L_2$).
    Hence, the objective $\Distance_\nu$ corresponds to minimizing the distance from a desired frequency vector $\nu$.
    \item For every $v \in V$, let $\kappa_v \subseteq [0,1]$ be an interval of admissible frequencies of visiting the vertex $v$. For example, if $\kappa_v = [0,0.2]$, then $v$ should be visited with frequency at most $0.2$.  For every $\mu \in \Dist(V)$, we put $\Satisfy_\kappa(\mu) = 0$ if $\mu(v) \in \kappa_v$ for all $v \in V$. Otherwise, $\Satisfy_\kappa(\mu) = 1$. The objective $\Satisfy_\kappa$ then corresponds to satisfying the constraints imposed by $\kappa$. 
\end{itemize}

In some scenarios, the value of a long-run average objective depends only on the total frequency of visits to ``equivalent'' vertices. Formally, such equivalence is defined as a \emph{labeling} $\calL\colon V \to L$ where equivalent vertices share the same label, and a \emph{labeled} long-run average objective is represented by a function  $\calL\textrm{-}\Obj\colon \Dist(L) \to \Rset^{\geq 0}$ specifying the ``badness'' of a given frequency of labels seen along a run. The function $\calL\textrm{-}\Obj$ represents the unique objective $\Obj\colon \Dist(V) \to \Rset^{\geq 0}$ such that $\Obj(\mu) =  \calL\textrm{-}\Obj(\mu_{\calL})$ where $\mu_{\calL}(\ell) = \sum_{v \in \calL^{-1}(\ell)} \mu(v)$.

In the following sections, we also apply $\Obj$ to distributions over augmented vertices $\ag{V}$. For every $\mu \in \Dist(\ag{V})$, we put $\Obj(\mu) = \Obj(\nu)$, where $\nu \in \Dist(V)$ is  defined by $\nu(v) = \sum_{m \in \alpha(v)} \mu(v,m)$.


\paragraph{Local Frequency Measures}
 
Let $D = (V,E,p)$ be an MDP and $\Obj$ a long-run average objective for $D$.

The ``global'' satisfaction of $\Obj$ achieved by an FR strategy  $\sigma$ is measured~by $\min_{B} \Obj(\Inv_B)$ where $B$ ranges over the BSCCs of $D^\sigma$. 
As we already noted in Example~\ref{exa-RM-system}, it may happen that an FR strategy achieves the optimal $\Obj(\Inv_B)$, but the expected value of $\Obj$ for a \emph{local} frequency of states sampled from $n$ consecutive states along a run is large. The \emph{local satisfaction} of $\Obj$ is measured by the \emph{expected badness of the local frequency} defined in the next paragraph.
 
Let $\sigma$ be a FR strategy, $B$ a BSCC of $D^\sigma$, and $\mu_B$ an initial distribution over~$B$. Consider the local frequency sampled from $n$ consecutive states along a run in~$B$, where the sampling starts in a randomly chosen \emph{pivot} state $p$. The probability of $p = s$ for a given $s \in B$ corresponds to the  ``global'' frequency of $s$ in a run, which is equal to $\Inv_B(s)$ independently of~$\mu_B$. Hence, the conditional expected badness of the local frequency under the condition $p = s$ is equal to $\Exp^{\mu_s}[\Obj(\Freq_n)]$ where $\mu_s$ is a distribution over $B$ such that $\mu_s(s) = 1$ and $\mu_s(t) = 0$ for $t \neq s$. Hence, the \emph{expected badness of the local frequency} is defined as
\[
     \sum_{s\in B} \Inv_B(s) \cdot \Exp^{\mu_s}[\Obj(\Freq_n)] \quad = \quad  \Exp^{\Inv_B}[\Obj(\Freq_n)]
\]



We intuitively expect that $\Exp^{\Inv_B}[\Obj(\Freq_n)]$ decreases with increasing time horizon~$n$. This holds if $n$ is increased by a \emph{sufficiently large~$k>0$}.  However, for $k=1$, it may happen that $\Exp^{\Inv_B}[\Obj(\Freq_n)]$ \emph{increases}. We fix this inconvenience by adopting the following definition: 
\[
  \LBad^\sigma(\Obj,d) \ = \ \min_B \min_{n\leq d} \Exp^{\Inv_B}[\Obj(\Freq_n)]    
\]
That is, for every $d \geq 1$, we consider the best outcome achievable for a time horizon of size \emph{at most} $d$ in a BSCC $B$ of $D^\sigma$. Note that $\LBad^\sigma(\Obj,d)$ is non-increasing in~$d$. 

The next theorem shows that the problem of computing an FR strategy $\sigma$ minimizing $\LBad^\sigma(\Obj,d)$ is computationally hard even for \emph{graphs} (MDPs with no stochastic vertices) where an optimal FR strategy does not require randomization. A proof is in \suppl.

\begin{theorem}
\label{thm-win-hardness}
    Let $D = (V,E,p)$ be a graph (i.e., $V_S = \emptyset$), $d \in \Nset$, and $\nu \in \Dist(V)$.
    The existence of a FR strategy $\sigma$ such that $\prob_{\Inv_B}[\Freq_n{=}\nu] = 1$ for some $n \leq d$ and a BSCC $B$ of $D^\sigma$ is NP-hard. 
    
    The NP-hardness holds even under the assumption that if such a $\sigma$ exists, it can be constructed so that $\sigma(\ag{v})$ is a Dirac distribution for every $\ag{v} \in \ag{V}$.
\end{theorem}

Note that Theorem~\ref{thm-win-hardness} implies NP-hardness of minimizing $\LBad^\sigma(\Obj,d)$ for $\Distance_\nu$ and $\Satisfy_\kappa$, because
$\prob_{\Inv_B}[\Freq_n{=}\nu] = 1$ iff 
\mbox{$\LBad^\sigma(\Distance_\nu,d) = 0$} iff 
$\LBad^\sigma(\Satisfy_\kappa,d) = 0$
where $\kappa(v) = [\nu(v),\nu(v)]$ for every $v \in V$.

%% file: sections/evaluate.tex
\section{Evaluating Local Badness}
\label{sec-eval}

In this section, we design algorithm \LE\ for evaluating $\LBad^\sigma(\Obj,d)$.

Let $D = (V,E,p)$ be an MDP, $\sigma$ an FR strategy for $D$, and $\calL\colon V \to L$ a labeling. Furthermore, let $\calL\textrm{-}\Obj{:}\break\Dist(L) \to \Rset^{\geq 0}$ be the desired objective function. Algorithm \LE\ consists of several phases, following the definition of $\LBad^\sigma(\Obj,d)$: 
First, we use Tarjan's algorithm \cite{Tarjan:SCC-decomp-SICOMP} to identify all BSCCs of $D^\sigma$.
For each BSCC $B$, the invariant distribution $\Inv_B$ is computed
via the following system of linear equations: For each $\ag{v}\in B$,
we have a fresh variable $z_{\ag{v}}$ and equations expressing that 
$z = z \cdot \Prob_B$ and $\sum_{\ag{v} \in B} z_{\ag{v}} = 1$. The vector $\Inv_B$ is the unique solution of this system. 

The core of \LE\ is Algorithm~\ref{alg:localeval} computing 
$\Exp^{\Inv_B}[\Obj(\Freq_n)\mid Init{=}\ag{v}]$ for all $\ag{v}\in B$ and $n\leq d$ by dynamic programming. Since for all $n\leq d$ we have that
\[
\Exp^{\Inv_B}[\Obj(\Freq_n)]=\sum_{\ag{v}\in B}\,\Inv_B(\ag{v})\cdot\Exp^{\Inv_B}[\Obj(\Freq_n)\mid Init{=}\ag{v}],
\]
the computation of $\LBad^\sigma(\Obj,d)$ is straightforward.

Algorithm \ref{alg:localeval} uses two associative arrays (e.g., C++ unordered\_map),
called $cur\_map$ and $next\_map$, to gather information about the probabilities of individual paths.
More specifically, the maps are indexed by \emph{states}, where a state consists of an augmented vertex $\ag{v}\in B$,
corresponding to the last vertex of a path, and a vector $vec$ of $|L|$ integers, corresponding
to the numbers of visits to particular labels. The value associated to a state $s$ is the total probability of
all paths corresponding to $s$. The values $\Exp^{\Inv_B}[\Obj(\Freq_n)\mid Init{=}\ag{v}]$ are gathered in a 2-dimensional array $\mathit{rsl}$.
Further details are given in \suppl.

\begin{algorithm}[t]
	\small
    \caption{The core procedure of \LE}
	\label{alg:localeval}
	\begin{algorithmic}
		\For{$\ag{v_0}\in B$}
            \State $s_0.\ag{v}$ = $\ag{v_0}$
            \State $s_0.vec$ = $\{0,\dots,0\}$
            \State $s_0.vec[\calL(v_0)]$++
			\State $cur\_map[s_0]$ = $1.$
			\For{$n\in\{1,\dots,d\}$}
				\For{$(s,p)\in cur\_map$}
					\State $rsl[\ag{v_0}][n]$ += $p\cdot\calL\textrm{-}\Obj(s.vec / n)$
				\EndFor
				\If{$n<d$}
					\For{$(s,p)\in cur\_map$}
						\For{$\ag{v}\in B$}
							\State $s'$ = $s$
							\State $s'.\ag{v}$ = $\ag{v}$
							\State $s'.vec[\calL(v)]$++
							\State $next\_map[s']$ += $p\cdot\sigma[s.\ag{v}][\ag{v}]$
						\EndFor
					\EndFor
					\State $swap(cur\_map,next\_map)$
					\State $next\_map.clear()$
				\EndIf
			\EndFor
		\EndFor
	\end{algorithmic}
\end{algorithm}

%% file: sections/optimize.tex
\section{Optimizing Local Badness}
\label{sec-optimize}

In this section, we design an algorithm \LS\ for constructing an FR strategy $\sigma$ with memory $\Mem$ minimizing $\LBad^\sigma(\Obj,d)$ for a given MDP~$D$. The main idea behind  \LS\ is to construct and optimize a function \emph{simultaneously} rewarding the following features of~$\sigma$:
\begin{description}
   \item[F1.] Global satisfaction of $\Obj$;
   \item[F2.] Stochastic stability of renewal times for families of augmented vertices with the same label.
   \item[F3.] The level of determinism achieved by $\sigma$.
\end{description}
Intuitively, F1 ensures that $\sigma$ achieves $\Obj$ globally, and F2 in combination with F3 ``encourage'' the features of $\sigma$ causing a small difference between the global and the local satisfaction. To understand the significance of F2, realize that the frequency of visits to augmented vertices with the same label is the inverse of the expected \emph{renewal time} for this family. Hence, the local stability of the frequency of visits can be achieved by maximizing the stochastic stability (i.e., minimizing the standard deviation of) the renewal time. To understand the significance of F3, realize that for every \emph{deterministic} FR strategy $\sigma$, the value of $\LBad^\sigma(\Obj,d)$ is equal to $\min_B \Obj(\Inv_B)$ for a \emph{sufficiently large}~$d$. Hence, putting more emphasis on F3 yields strategies where $\sigma(\ag{v})$ is close to a Dirac distribution for many $\ag{v}$, which may be advantageous when~$d$ is high.

\subsection{Measuring F1--F3}
\label{sec-F-measures}

In this section, we design efficient measures for F1--F3 and combine these measures into a single function $\Comb$.

Let $D = (V,E,p)$ be an MDP, $\calL\colon V \to L$ a labeling, and $\Obj$ a long-run average objective for $D$. Furthermore, let $\sigma$ be an FR strategy with memory $\Mem$ and $B$ a BSCC of $D^\sigma$. Recall that $\Inv_B$ is the invariant distribution of $B$, and $B^{\Inv_B}$ is the Markov chain determined by $B$ and the initial distribution $\Inv_B$. Furthermore, let $\RT(w)$ be the least \mbox{$i\geq 1$} such that $\calL(\ag{v}_i)  = \calL(\ag{v}_0)$. If there is no such $i$, we put $\RT(w) = \infty$. Hence, $\RT(w)$ is the number of edges needed to visit an augmented vertex with the same label as $\Init(w)$ (i.e., the Renewal Time to the initial label).

\paragraph{Measuring F1} The global \emph{dissatisfaction} of $\Obj$ achieved by $\sigma$ in $B^{\Inv_B}$
is measured by $\Obj(\Inv_B)$.

\paragraph{Measuring F2} 
%
Let $\Var^{\Inv_B}[\RT \mid \calL(\Init) {=} \ell]$ be the conditional variance of the Renewal Time to the initial label under the condition that a run is initiated in an augmented vertex with label $\ell$. If the probability of $\calL(\Init) {=} \ell$ is zero, i.e, $\Inv_B$ assigns zero to all augmented vertices with label $\ell$, we treat $\Var^{\Inv_B}[\RT \mid \calL(\Init) {=} \ell]$ as zero.
Furthermore, we define the corresponding standard deviation
\[
   \SD(\ell) = \sqrt{\Var^{\Inv_B}[\RT \mid \calL(\Init) {=} \ell]} \,.  
\]
Stochastic \emph{instability} of renewal times caused by $\sigma$ in $B^{\Inv_B}$ is measured by the function 
\[
   \Penalty_1(\sigma,B) =  \sum_{\ell \in L} \Inv_B(\ell) \cdot \SD(\ell)   
\]
where $\Inv_B(\ell)$ is the sum of all $\Inv_B(\ag{v})$ where $\ag{v} \in B$ and $\calL(v) = \ell$.
That is, $\Penalty_1$ is the weighted sum of all $\SD(\ell)$ where the weights correspond to the limit label frequencies.

\paragraph{Measuring F3} The level of \emph{non-determinism} caused by $\sigma$ in $B^{\Inv_B}$ is measured by the stochastic instability of Renewal Times separately for each augmented vertex. That is, we put 
\[
   \Penalty_2(\sigma,B) =  \sum_{\ag{v} \in B} \Inv_{B}(\ag{v}) \cdot \sqrt{\Var^{\Inv_B}[\RT \mid \Init {=} \ag{v}]}   
\]
If the probability of $\Init {=} \ag{v}$ is zero, we treat the corresponding conditional variance as zero. 

Note that for every \emph{deterministic} strategy we have that $\Var^{\Inv_B}[\RT \mid \Init {=} \ag{v}] = 0$ for every $\ag{v}$, i.e., $\Penalty_2 = 0$. However, $\Penalty_1$ is still positive if the expected renewal times for the individual augmented vertices with the same label differ. The only ``degenerated'' case when $\Penalty_1$ and $\Penalty_2$ are the same functions is when all vertices have pairwise different labels and every vertex is allocated just one memory state.

\paragraph{Combining the measures}
Our \LS\ algorithm attempts to \emph{minimize} the following function $\Comb(\sigma)$ over all BSCC~$B$ of $D^\sigma$:
{\small\[
     (1{-}\beta{-}\gamma) \Obj(\Inv_B) + \beta  \cdot c_1 \cdot  \Penalty_1(\sigma,B) + \gamma \cdot c_2 \Penalty_2(\sigma,B)
\]}%
where  $\beta,\gamma \in [0,1]$ are \emph{weights} such that $\beta+\gamma < 1$ representing the preference among F1--F3. Since the values of $\Obj(\Inv_B)$ may range over very different intervals than $\Penalty_1$ and $\Penalty_2$, we also use the \emph{normalizing constants}
$c_1 = (\Obj(\Inv_B){+}1)/(\Penalty_1(\sigma,B){+}1)$ and 
$c_2 = (\Obj(\Inv_B){+}1)/(\Penalty_2(\sigma,B){+}1)$.

\begin{algorithm}[t]
   \small
   \caption{\LS}
   \label{alg:optim}
   \begin{algorithmic}[t]
   \State ${\rm SolutionParameters}\gets {\it RandomInit}$
   \For{$i\in \{1,\ldots,{\rm Steps}\}$} 
           \State ${\sigma} \gets {\it Softmax}({\rm SolutionParameters})$
           \State ${\Comb(\sigma)} \gets {\it EvaluateComb}({\sigma})$
           \State ${\nabla \Comb(\sigma)} \gets {\it Gradient}({\sigma})$
           \State ${\rm SolutionParameters~~} {+}{=} {\it ~~Step}({\nabla \Comb(\sigma)})$
           \State \textbf{Save} ${\Comb(\sigma)}, {\sigma}$\vspace{0.5ex}
   \EndFor
   \Return ${\sigma}$ with the least $\Comb(\sigma)$
   \end{algorithmic}
\end{algorithm}

\begin{figure*}
    \includegraphics*[width=.55\textwidth]{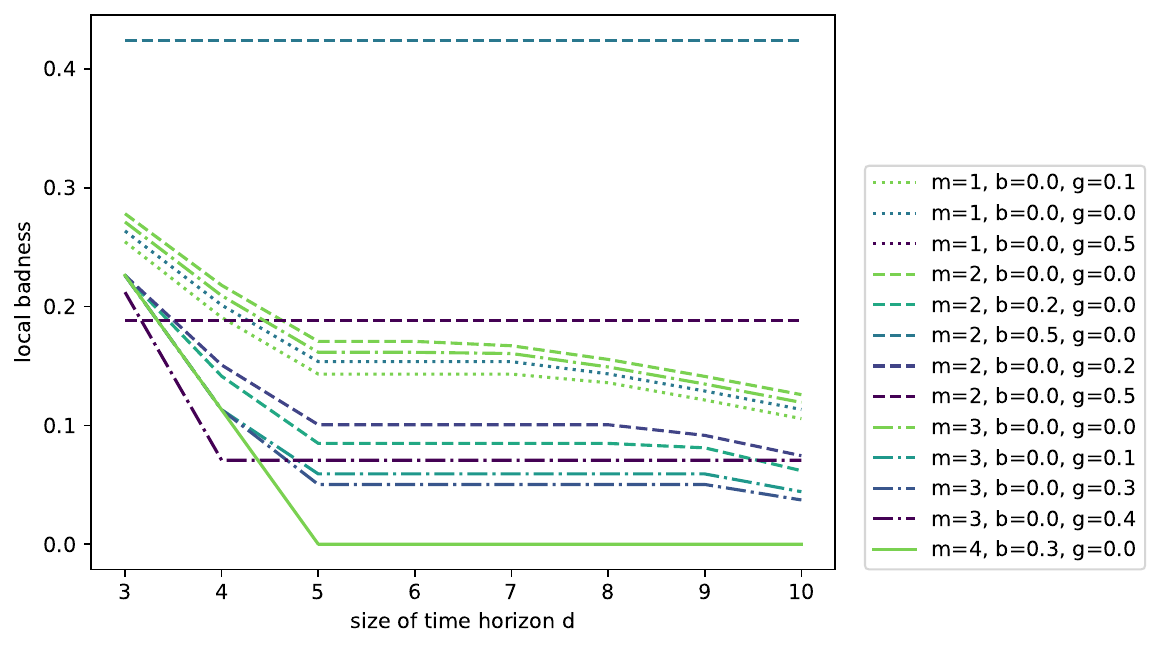}
    \includegraphics*[width=.45\textwidth]{l-badness_at_5_a_memory=3.pdf}
    \caption{Strategies constructed for the graph of Fig.~\ref{fig-RM-system}. Adding more memory to the state $R$ helps. Best results are achieved for certain combinations of $\beta$ and $\gamma$ where $\beta+\gamma$ does not exceed the threshold around $0.6$.}
    \label{fig-exp-I}
\end{figure*}

\subsection{Computing $\Comb$}
\label{sec-eval-Comb}
In this section, we show that there exist three efficiently constructible systems of linear equations with unique solutions $\vec{x}$, $\vec{y}$ and $\vec{z}$ such that the function $\Comb$ is a closed-form expression over the components of $\vec{x}$, $\vec{y}$, and $\vec{z}$ containing only differentiable functions. 
This allows us to compute the \emph{gradient} of $\Comb$ efficiently and apply state-of-the-art methods of differentiable programming to minimize $\Comb$ by gradient descent, which is the essence of \LS\ functionality. 

For every $\ag{v} \in B$ and $\ell \in L$ such that $\Inv_B(\ell) > 0$, let $x_{\ag{v},\ell}$ and 
$y_{\ag{v},\ell}$ be fresh variables. For every $x_{\ag{v},\ell}$, we add an equation
\[
   x_{\ag{v},\ell} = 
   \begin{cases}
      0 & \mbox{if $\calL(v) = \ell$,}\\
      1 + \sum_{\ag{u} \in B} \sigma(\ag{v})(\ag{u}) \cdot x_{\ag{u},\ell}  & \mbox{otherwise.}
   \end{cases}
\]
Then the system has a unique solution $\vec{x}$ where $\vec{x}_{\ag{v},\ell}$ is the expected time for visiting an $\ell$-labeled augmented vertex from $\ag{v}$. Hence, 
$\Exp^{\Inv_B}[RT \mid \Init = \ag{v}] = 1 + \sum_{\ag{u} \in B} \sigma(\ag{v})(\ag{u}) \cdot \vec{x}_{\ag{u},\ell}$, where $\ell = \calL(v)$.

Similarly, for every $y_{\ag{v},\ell}$, we add an equation
\[
   y_{\ag{v},\ell} = 
   \begin{cases}
      0 & \mbox{if $\calL(v) = \ell$,}\\
      1 + \sum_{\ag{u} \in B} \sigma(\ag{v})(\ag{u}) \cdot (2\vec{x}_{\ag{u},\ell} + y_{\ag{u},\ell})  & \mbox{otherwise.}
   \end{cases}
\]
Note that the above equation is \emph{linear} and uses components of $\vec{x}$ in the coefficients. The system has a unique solution $\vec{y}$ where $\vec{y}_{\ag{v},\ell}$ is the expected \emph{square} of the time for visiting an $\ell$-labeled augmented vertex from $\ag{v}$. Hence, 
\[
   \Exp^{\Inv_B}[RT^2 \mid \Init = \ag{v}] = 1 + \sum_{\ag{u} \in B} \sigma(\ag{v})(\ag{u}) \cdot (2\vec{x}_{\ag{u},\ell} + \vec{y}_{\ag{u},\ell})
\] 
The vector $\vec{z}$ corresponding to the invariant distribution $\Inv_B$ is computed the same way as in \LE\ (Section \ref{sec-eval}).

Both $\Exp^{\Inv_B}[RT \mid \calL(\Init) {=} \ell]$ and $\Exp^{\Inv_B}[RT^2 \mid \calL(\Init) {=} \ell]$ are weighted sums of $\Exp^{\Inv_B}[RT \mid \Init = \ag{v}]$ and $\Exp^{\Inv_B}[RT^2 \mid \Init = \ag{v}]$ where the weights are expressions over the components of $\vec{z}$. Since 
$\Var[X \mid Y] = \Exp[X^2 \mid Y] - \Exp^2[X \mid Y]$ for all random variables $X,Y$, the
conditional variances $ \Var^{\Inv_B}[\RT \mid \Init {=} \ag{v}]$ and $\Var^{\Inv_B}[\RT \mid \calL(\Init) {=} \ell]$ are also expressible as closed form expressions over $\vec{x}$, $\vec{y}$, and $\vec{z}$. Hence, $\Comb$ also has this property.

\subsection{The \LS\ Algorithm}
\label{sec-ls-alg}

Our algorithm is based on differentiable programming and gradient descent, and it performs the standard optimization loop shown in Algorithm~\ref{alg:optim}. For every pair of augmented vertices $(\ag{v},\ag{u})$ such that $(v,u) \in E$, we need a parameter representing $\sigma(\ag{v})(\ag{u})$. Note that if $v$ is stochastic, then the parameter actually represents the probability of selecting the memory state of $\ag{u}$. These parameters are initialized to random values sampled from \textit{LogUniform} distribution (so that we impose no prior knowledge about the solution). Then, they are transformed into probability distributions using the standard \textit{Softmax} function.
 
The crucial ingredient of \LS\ is the procedure \textit{EvaluateComb} for computing the value of $\Comb$ for the strategy represented by the parameters (see Section~\ref{sec-eval-Comb}).
 This procedure allows to compute $\Comb(\sigma)$, and also the gradient of $\Comb(\sigma)$ at the point corresponding to $\sigma$ by automatic differentiation. After that, we update the point representing the current $\sigma$ in the direction of the steepest descent. The intermediate solutions and the corresponding $\Comb$ values are stored, and the best solution found within ${\rm Steps}$ optimization steps is returned. Our implementation uses
\textsc{PyTorch} framework~\cite{PyTorch} and its automatic differentiation with \textsc{Adam} optimizer~\cite{Adam}). 

Observe that \LS\ is equally efficient for general MDPs and graphs. The only difference is that stochastic vertices generate fewer parameters.

%% file: sections/experiments.tex
\section{Experiments}
\label{sec-experiments}
The system setup was as follows: CPU: AMD Ryzen 93900X (12 cores); RAM: 32GB; Ubuntu 20.04. To separate the probabilistic choice introduced by the constructed strategies from the internal probabilistic choice performed in stochastic vertices, we perform our experiments on graphs. 


\subsection{Experiment~I}
In our first experiment, we aim to analyze the impact of the $\beta,\gamma$ coefficients in $\Comb$ and the size of available memory on the structure and performance of the resulting strategy $\sigma$. 

We use the graph $D$ of Fig.~\ref{fig-RM-system}(a) and the objective $\Distance_\nu$ with $L_2$ norm where $\nu(R) = \frac{4}{5}$ and $\nu(M) = \frac{1}{5}$. 
In our FR strategies, we allocate $m \leq 4$ memory states to the vertex~$R$ and one memory state to the vertex~$M$. The coefficients $\beta,\gamma$ range over $[0, 0.5]$ with a discrete step $0.1$. For every choice of $\beta$, $\gamma$, and $m$, we run  \LS\  $40$ times with ${\rm Steps}$ set to $800$ and return the strategy $\sigma$ with the \emph{least} value of $\Comb$ found. Then, we use the \LE\ algorithm to compute $\LBad^\sigma(\Distance_\nu,d)$ for $d \in \{3,\ldots,10\}$.

\paragraph{Discussion}
The plot of Fig.~\ref{fig-exp-I}~(left) shows that 
\begin{itemize}
    \item[1.] increasing the size of memory $m$ leads to better performance (smaller $\LBad^\sigma(\Distance_\nu,d)$);
    \item[2.] setting $\beta {=} \gamma {=} 0$ produces worse strategies (for every~$m$) than setups with even small positive values of $\beta,\gamma$;
    \item[3.] setting $\beta + \gamma \geq 0.5$ leads to very bad strategies.
\end{itemize}
The outcomes 1.{} and 2.{} are in full accordance with the intuition presented in Section~\ref{sec-optimize}. Outcome~3.{} is also easy to explain---when $\beta$ or $\gamma$ is too large, the algorithm \LS\ concentrates on maximizing stochastic stability of renewal times or achieving determinism and ``ignores'' the $L_2$ distance from $\nu$. For example, the worst strategy of Fig.~\ref{fig-exp-I}~(left) obtained for $m=2$, $\beta=0.5$, $\gamma = 0$ ``regularly alternates'' between $R$ and $M$, i.e., the renewal times of $R$ and $M$ are equal to~$2$ and have \emph{zero variance}. This leads to local frequency $(\frac{1}{2},\frac{1}{2})$, which is ``far'' from the desired~$\nu$.  

We also provide plots of $\LBad^\sigma(\Distance_\nu,d)$ where $m$ and $d$ are fixed and $\beta,\gamma$ range over $[0, 0.5]$. The plot for $d=5$ and $m=3$ is shown in Fig.~\ref{fig-exp-I}~(right). All these plots (see \suppl) consistently show that the best outcomes are achieved for certain combinations of $\beta$ and $\gamma$ where $\beta+ \gamma$ is positive but below $0.6$ (in Fig.~\ref{fig-exp-I}~(right), the best outcomes are in yellow). 

\subsection{Experiment~II}

\begin{figure}[t]\centering
    \begin{tikzpicture}[x=4cm, y=2.5cm, >=stealth', scale=0.48,font=\small]
        \coordinate (a) at (-1.1,-2); 
        \node (l) at ($(a) +(-.4,0)$) {$(a)$};
        \foreach \x/\y/\m in {1/0/1, 2/0.8/2, 3/1.6/2, 4/2.4/2}{%
            \node[min] (N\x) at ($(a) +(\y,0)$) {$v_{\x}$};
            \node (L\x) at ($(a) +(\y,0) +(.2,.2)$) {\color{brown}{$\frac{\x}{10}$}};
            \node (M\x) at ($(a) +(\y,0) +(.2,-.2)$) {\color{red}{$\m$}};
         }
        \foreach \x/\y in {1/2, 2/3, 3/4}{%
           \draw[->,thick] (N\x) to  (N\y);
        }
        \draw[->,thick, rounded corners] (N4) -- ($(N4) +(0,-.45)$) -| (N1);
        \foreach \x in {1, 2, 3, 4}{%
            \path (N\x) edge [out=90,in=140,thick, looseness=8, ->]  (N\x);
        }
        \coordinate (a) at (-1.1,-3.2); 
        \node (l) at ($(a) +(-.4,0)$) {$(b)$};
        \foreach \x/\y/\i in {1/0/1, 2/0.6/2, 3/1.2/2, 4/1.8/3, 5/2.4/3, 6/3/4, 7/3.6/4}{%
            \node[min] (N\x) at ($(a) +(\y,0)$) {$v_{\i}$};
         }
        \foreach \x/\y in {1/2, 2/3, 3/4, 4/5, 6/7}{%
           \draw[->,thick] (N\x) to  (N\y);
        }
        \draw[->,thick, rounded corners] (N7) -- node[right] {$\frac{1}{3}$} ($(N7) +(0,-.45)$) -| (N1);
        \draw[->,thick] (N5) -- node[above] {$\frac{1}{2}$} (N6);
        \path (N5) edge [out=90,in=140,thick, looseness=8, ->] node[right=0.5em] {$\frac{1}{2}$} (N5);
        \path (N7) edge [out=90,in=140,thick, looseness=8, ->] node[right=0.5em] {$\frac{2}{3}$} (N7);          
        \coordinate (a) at (-1.1,-4.4); 
        \node (l) at ($(a) +(-.4,0)$) {$(c)$};
        \foreach \x/\y/\i in {1/0/1, 2/0.6/2, 3/1.2/2, 4/1.8/3, 5/2.4/3, 6/3/4, 7/3.6/4}{%
            \node[min] (N\x) at ($(a) +(\y,0)$) {$v_{\i}$};
         }
        \foreach \x/\y in {1/2, 2/3, 3/4}{%
           \draw[->,thick] (N\x) to  (N\y);
        }
        \draw[->,thick, rounded corners] (N7) -- node[right] {$\frac{1}{2}$} ($(N7) +(0,-.45)$) -| (N1);
        \draw[->,thick] (N4) -- node[above, near start] {$\frac{2}{3}$} (N5);
        \draw[->,thick] (N5) -- node[above, near start] {$\frac{2}{3}$} (N6);
        \draw[->,thick] (N6) -- node[above, near start] {$\frac{1}{2}$} (N7);
        \path (N4) edge [out=90,in=140,thick, looseness=8, ->] node[right=0.5em] {$\frac{1}{3}$} (N4);
        \path (N5) edge [out=90,in=140,thick, looseness=8, ->] node[right=0.5em] {$\frac{1}{3}$} (N5);
        \path (N6) edge [out=90,in=140,thick, looseness=8, ->] node[right=0.5em] {$\frac{1}{2}$} (N6); \path (N7) edge [out=90,in=140,thick, looseness=8, ->] node[right=0.5em] {$\frac{1}{2}$}(N7);       
    \end{tikzpicture}
    \caption{The structure of $D_4$ $(a)$, $\pi_4$ $(b)$, and $\varrho_4$~$(c)$.}
    \label{fig-Dn-instance}
    \end{figure}
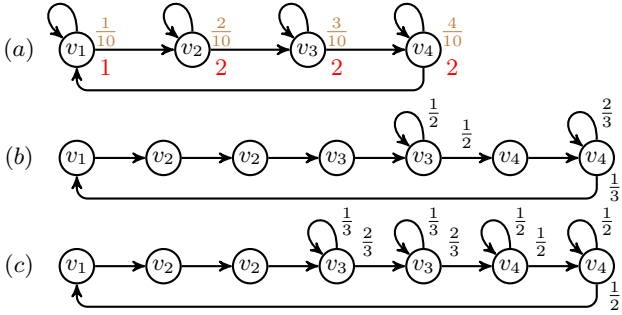

Here we aim to analyze the scalability of \LE\ and \LS\ and demonstrate that \LS\ can produce sophisticated strategies for instances of non-trivial size. Since the running time of \LS\ depends on the \emph{number of parameters}, i.e., the number of augmented edges, we need to consider a scalable instance.

For every $n\geq 2$, let $D_n$ be a graph with vertices $v_1,\ldots,v_n$ and edges $(v_i,v_i)$ and $(v_i,v_{(i\bmod n)+1})$ for every $i \leq n$. Every $v_i$ is assigned $\min\{i,\lceil \frac{n}{2} \rceil\}$ memory states.
The desired frequency $\nu$ is defined by $\nu(v_i) = i/s$ where $s = \frac{n(n+1)}{2}$. The structure of $D_4$ is shown in Fig.~\ref{fig-Dn-instance}(a), together with $\nu$ (brown) and memory allocation (red).

We consider the objective $\Distance_\nu$ with the $L_2$ norm, and we aim to optimize $\LBad^\sigma(\Distance_\nu,d)$ where $d = s$ (the least $d$ such that $\nu$ is achievable in $d$ consecutive states.) To evaluate the scalability of \LE\ and \LS\, we run \LE\ $5$ times for different choice of $\beta$ and $\gamma$ and evaluate $\LBad^\sigma(\Distance_\nu,d)$ using \LE\ and also a naive algorithm based on depth-first search (see \suppl\ for a more detailed description of the naive algorithm). In Table~\ref{tab-times}, for every $n$ we report the number of parameters, the size of $d$, the average time of one Step of \LS\ (i.e., one iteration of the main \textbf{for} loop of \LS), one run of \LE, and one run of the naive evaluation algorithm (in secs). 

To evaluate the quality of strategies constructed by \LS, we consider two natural strategies $\pi_n$ and $\varrho_n$ (see Fig.~\ref{fig-Dn-instance}~(b) and~(c)). Both strategies perform an ``ideal'' number of self-loops on $v_1,\ldots,v_{\lceil n/2 \rceil}$ where the memory suffices. On the other vertices, $\pi_n$ performs $\lceil n/2 \rceil -1$ self-loops deterministically and then selects randomly between the self-loop and the edge to the next vertex, while $\varrho_n$ performs a random choice in every visit. The probabilities are computed so that $\Inv = \nu$. Hence, both $\pi_n$ and $\varrho_n$ represent an ``educated guess'' for a high-quality strategy. 

For all $n \in \{2,\ldots,8\}$, we run \LS\ $40$ times with Steps set to $800$ for all $\beta,\gamma \in \langle0, 0.5\rangle$ with a discrete step $0.1$, always collecting the strategy $\sigma_n$ with the minimal $\Comb$ value. The outcomes are shown in Table~\ref{tab-synt}. Interestingly, $\sigma_n$ \emph{significantly outperforms} both $\pi_n$ and $\varrho_n$ for all $n\geq 4$. The strategy $\sigma_n$ cannot be found ``ad-hoc''; in most cases, the associated invariant distribution is different from $\nu$, which means that global satisfaction ``traded'' for local satisfaction. We also report the values of $\beta$ and $\gamma$ for which the best strategy $\sigma_n$ was found by \LS.

\paragraph{Discussion} Table~\ref{tab-times} shows that \LS\ can easily process instances with thousands of parameters, while the scalability limits of \LE\ are reached for $d \approx 30$. Hence, \LE\ cannot be used for strategy synthesis based on gradient descent because \LE\ would have to be invoked hundreds of times in a single run. Table~\ref{tab-synt} shows that \LS\ can construct sophisticated strategies for non-trivial instances. Details are in \suppl. 




\begin{table}[t]
    \footnotesize
    \centering
    \begin{tabular}{cccccc}
    \toprule     
    & \multicolumn{5}{c}{$\LBad(\Distance_\nu,d)$}\\
     $n$ & $\pi_n$ & $\varrho_n$ & $\sigma_n$ & $\beta$ & $\gamma$\\ 
     \cmidrule(r){1-6}
     $2$  & $0.15713$ & $0.15713$ & $0.15713$  & $0.2$ & $0.0$\\
     $3$  & $0.11479$ & $0.10255$ & $0.11473$  & $0.1$ & $0.1$\\
     $4$  & $0.19416$ & $0.17131$ & $0.10540$  & $0.0$ & $0.2$\\ 
     $5$  & $0.14277$ & $0.11762$ & $0.10540$  & $0.0$ & $0.2$\\
     $6$  & $0.17491$ & $0.13985$ & $0.08016$  & $0.0$ & $0.2$\\
     $7$  & $0.13781$ & $0.10456$ & $0.10022$  & $0.0$ & $0.2$\\
     $8$  & $0.15609$ & $0.11436$ & $0.10012$  & $0.0$ & $0.2$\\
    \bottomrule
    \end{tabular}
    \caption{Strategy $\sigma_n$ outperforms $\pi_n$ and $\varrho_n$.}
    \label{tab-synt}
\end{table}

\begin{table}[t]
    \footnotesize
    \centering
    \begin{tabular}{cccccc}
    \toprule
     $n$ & \textit{Par} & $d$ & \textit{Step} & \LE & \textit{Naive}\\ 
     \cmidrule(r){1-6}
     $4$  & $25$ & $10$ & 2.12E-03 & 2.21E-04  & 2.74E-03 \\
     $5$  & $61$ & $15$ & 2.71E-03 & 4.56E-03  & 1.21E+02\\
     $6$  & $79$ & $21$ & 2.21E-03 & 4.13E-01  & timeout  \\
     $7$  &  $150$ & $28$ & 2.44E-03 & 1.98E+01   & timeout\\
     $8$  & $182$ & $36$ & 2.50E-03 & timeout &  timeout \\
     $10$  & $350$ & $55$ & 2.97E-03 & timeout &  timeout \\
     $12$  & $599$ & $78$ & 6.43E-03 & timeout & timeout  \\
    $14$  & $945$ & $105$ & 1.88E-02 & timeout &  timeout\\
    $16$  & $1404$ &  $136$ & 3.91E-02& timeout &  timeout \\
     $18$  & $1992$ & $171$ & 1.05E-01& timeout &  timeout \\
     $20$  & $2725$ &  $210$ & 2.15E-01 & timeout & timeout \\
    \bottomrule
    \end{tabular}
    \caption{Running times in \emph{seconds}, timeout $= 900$ secs.}
    \label{tab-times}
\end{table}


    \section{Conclusions} 
    The results demonstrate that non-trivial instances of the local satisfaction problem for long-run average objectives can be solved efficiently despite the \mbox{NP-hardness} of this problem. Experiment~II also shows that the best strategy for $D_n$ is obtained by setting $\beta=0.0$ and $\gamma=0.2$. Although \LE\ cannot evaluate the strategies obtained for large $n$'s, there is a good chance that these strategies are better than the ones constructed ad-hoc. This indicates how to overcome the scalability issues for other parameterized instances. 
    

%% file: suppsections/hardness.tex
\newcommand{\Cl}{\mathit{Cl}}
\newcommand{\len}{\mathit{length}}
\renewcommand{\Var}{\mathit{Var}}

\section{Hardness of the Local Satisfaction Problem}
\label{supsec-hardness}

Let us recall Theorem~1 presented in Section~2 in the main body of the paper.

\begin{theorem}
    Let $D = (V,E,p)$ be a graph (i.e., $V_S = \emptyset$), $d \in \Nset$, and $\nu \in \Dist(V)$.
    The existence of a FR strategy $\sigma$ such that $\prob_{\Inv_B}[\Freq_n{=}\nu] = 1$ for some $n \leq d$ and a BSCC $B$ of $D^\sigma$ is NP-hard. 
        
    The NP-hardness holds even under the assumption that if such a $\sigma$ exists, it can be constructed so that $\sigma(\ag{v})$ is a Dirac distribution for every $\ag{v} \in \ag{V}$.
\end{theorem}

Proof follows easily by reducing the NP-complete Hamiltonian Cycle (HC) problem. An instance of HP is a directed graph $D = (V,E)$, and the question is whether there exist a \emph{Hamiltonian cycle} visiting each vertex precisely once. Let $\nu \Dist(V)$ such that $\nu(v) = 1/|V|$ for every $v\in V$, and let $d = |V|$. 

If the graph $D$ contains a Hamiltonian cycle, then the cycle can be represented as a \emph{memoryless} strategy $\sigma$ such that  $\prob_{\Inv_B}[\Freq_d{=}\nu] = 1$. Conversely, let $\sigma$ be a FR strategy for $D$ such that $\prob_{\Inv_B}[\Freq_n{=}\nu] > 0$ for some $n \leq d$ and a BSCC of $D^\sigma$. Then $n=d$, because $\nu$ is not achievable in a strictly shorter time horizon. Furthermore, for every $\ag{v} \in B$ we have that $\prob_{\Inv_B}[\Freq_d {=}\nu \mid \Init{=}\ag{v}] = 1$. Let $w = \ag{v}_0,\ag{v}_1,\ldots$ be a run in $B$ such that $\sigma(\ag{v}_i)(\ag{v}_{i+1}) > 0$ for every $i \geq 0$, and let $w_1 = \ag{v}_1,\ag{v}_2,\ldots$ be the run obtained from $w$ by ``cutting off'' the initial augmented vertex~$\ag{v}_0$. Then $\Freq_d(w) = \Freq_d(w_1) = \nu$. This means that both $w$ and $w_1$ must visit an augmented vertex of the form $\ag{u}$ for every $u \in V$ in the first $d$ states, which is possible only if \emph{precisely one} such $\ag{u}$ is visited. This implies that the sequence $\ag{v}_0,\ldots,\ag{v}_d$ represents a Hamiltonian cycle in~$D$. 

Hence, $D$ contains a Hamiltonian cycle iff there is a FR strategy $\sigma$ such that $\prob_{\Inv_B}[\Freq_n{=}\nu] = 1$ for some $n \leq d$ and a BSCC $B$ of $D^\sigma$. Furthermore, if such a $\sigma$ exists, it can be constructed so that it is \emph{memoryless} and  $\sigma(\ag{v})$ is a Dirac distribution for every $\ag{v} \in \ag{V}$.

A similar proof shows that the existence of a FR strategy $\sigma$ such that $\prob_{\Inv_B}[\Freq_n{=}\nu] > 0$ for some $n \leq d$ and a BSCC $B$ of $D^\sigma$ is also NP-hard (here we reduce the Hamiltonian \emph{path} problem which is also NP-hard). Again, this holds even under the assumption that if such a $\sigma$ exists, it can be constructed to that it is memoryless and $\sigma(\ag{v})$ is a Dirac distribution for every $\ag{v} \in \ag{V}$.

Let us note that for MDPs with stochastic vertices, the local satisfaction problem is even \PSPACE-hard. This follows by reducing the cost problem for acyclic cost process, which is \PSPACE-complete \cite{HK:staying-on-budget}.

%% file: suppsections/algorithms.tex
\section{Evaluation Algorithms}
In this section, we provide a pseudocode of the naive \mbox{DFS-based} evaluation algorithm
which was used in Experiment~II (column \textit{Naive} of Table 2). We also present further details regarding the core procedure of our \LE\ algorithm.

\subsection{Naive Algorithm}
\begin{algorithm}[h]
	\small
    \caption{DFS procedure of the naive evaluation algorithm}
	\label{alg:DFS}
	\begin{algorithmic}[1]
		\State $rsl[\ag{v_0}][n]$ += $p\cdot\calL\textrm{-}\Obj(vec / n)$
		\If{$n<d$}
			\For{$\ag{u}\in B$}
                \State $vec[\calL(u)]$++
                \State DFS($\ag{v_0},\ag{u},p\cdot\sigma[\ag{v}][\ag{u}],n+1,vec$)
                \State $vec[\calL(u)]$--$\,$--
            \EndFor
		\EndIf
	\end{algorithmic}
\end{algorithm}
The DFS procedure inputs the following parameters:
\begin{itemize}
    \item the initial augmented vertex $\ag{v_0}$;
    \item the current augmented vertex $\ag{v}$;
    \item the probability $p$ of the current path;
    \item the length $n$ of the current path;
    \item the vector $vec$ with the same meaning as in \LE.
\end{itemize}
The procedure is called as DFS($\ag{v_0},\ag{v_0},1.,1,vec$)
for each $\ag{v_0}$ in the currently examined BSCC $B$,
where $vec=\{0,0,\dots,1,\dots,0,0\}$ with the 1 at position $\calL(v_0)$.
The output is the same as in the core procedure of \LE, \ie,
at the end of the computation, $rsl[\ag{v_0}][n]$ is equal to
$\Exp^{\Inv_B}[\Obj(\Freq_n)\mid Init{=}\ag{v}]$ for each $\ag{v_0}\in B$ and $n\leq d$.

\subsection{\LE\ Details}
Let us recall the core procedure of \LE:
\begin{algorithm}[h]
	\small
    \caption{The core procedure of \LE}
	\label{alg:localeval}
	\begin{algorithmic}[1]
		\For{$\ag{v_0}\in B$}
            \State $s_0.\ag{v}$ = $\ag{v_0}$
            \State $s_0.vec$ = $\{0,\dots,0\}$
            \State $s_0.vec[\calL(v_0)]$++
			\State $cur\_map[s_0]$ = $1.$
			\For{$n\in\{1,\dots,d\}$}
				\For{$(s,p)\in cur\_map$}
					\State $rsl[\ag{v_0}][n]$ += $p\cdot\calL\textrm{-}\Obj(s.vec / n)$
				\EndFor
				\If{$n<d$}
					\For{$(s,p)\in cur\_map$}
						\For{$\ag{v}\in B$}
							\State $s'$ = $s$
							\State $s'.\ag{v}$ = $\ag{v}$
							\State $s'.vec[\calL(v)]$++
							\State $next\_map[s']$ += $p\cdot\sigma[s.\ag{v}][\ag{v}]$
						\EndFor
					\EndFor
					\State $swap(cur\_map,next\_map)$
					\State $next\_map.clear()$
				\EndIf
			\EndFor
		\EndFor
	\end{algorithmic}
\end{algorithm}

The labels and augmented vertices are represented by the least $|L|$ and $|\ag{V}|$, respectively,
non-negative integers, so that they can be used directly as indices into arrays. In order to use
states $(s,vec)$ as indices into the maps (which are implemented as C++ unordered\_map), one has
to define a hashing function. We use the function
\[
    (s,vec)\mapsto h(s \oplus \bigoplus_{\ell\in L}\,vec[\ell] << (3(\ell+1)\bmod 64))
\]
where $\oplus$ denotes the bitwise xor, $<<$ denotes the left shift, and $h$ is the standard C++ hashing function for integers.

Finally, we mention some optimizations which are omitted in the pseudocode for ease of presentation:
The swap at line 16 can be realized trivially by swapping pointers.
The loop at line 11 (and at line 3 in the DFS procedure) can be sped up by precomputing a successor list
for each $\ag{v}$ (\ie, a list of those $\ag{u}$ for which $\sigma[\ag{v}][\ag{u}]>0$)
and iterating only through the successors.
These optimizations are included in our implementations.

%% file: suppsections/experiments.tex
\section{Detailed Experimental Results}

In this section, we report additional details about the experiments presented in Section~5 in the main body of the paper.

\subsection{Experiment~I}
\begin{figure*}
    \includegraphics*[width=.5\textwidth]{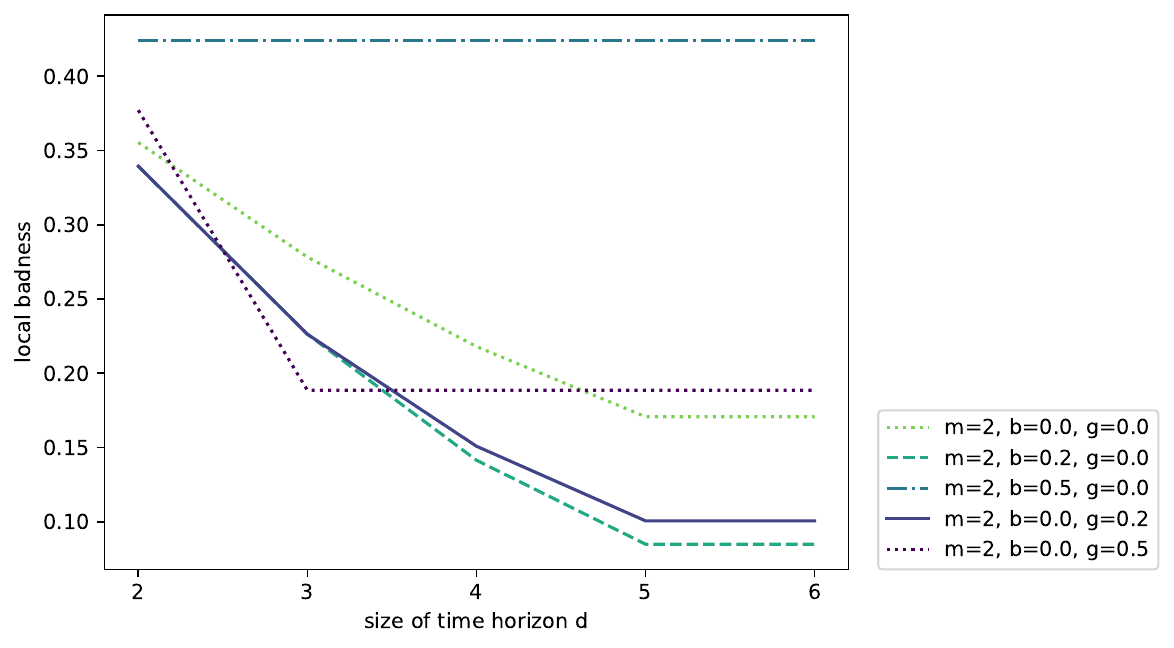}
    \includegraphics*[width=.5\textwidth]{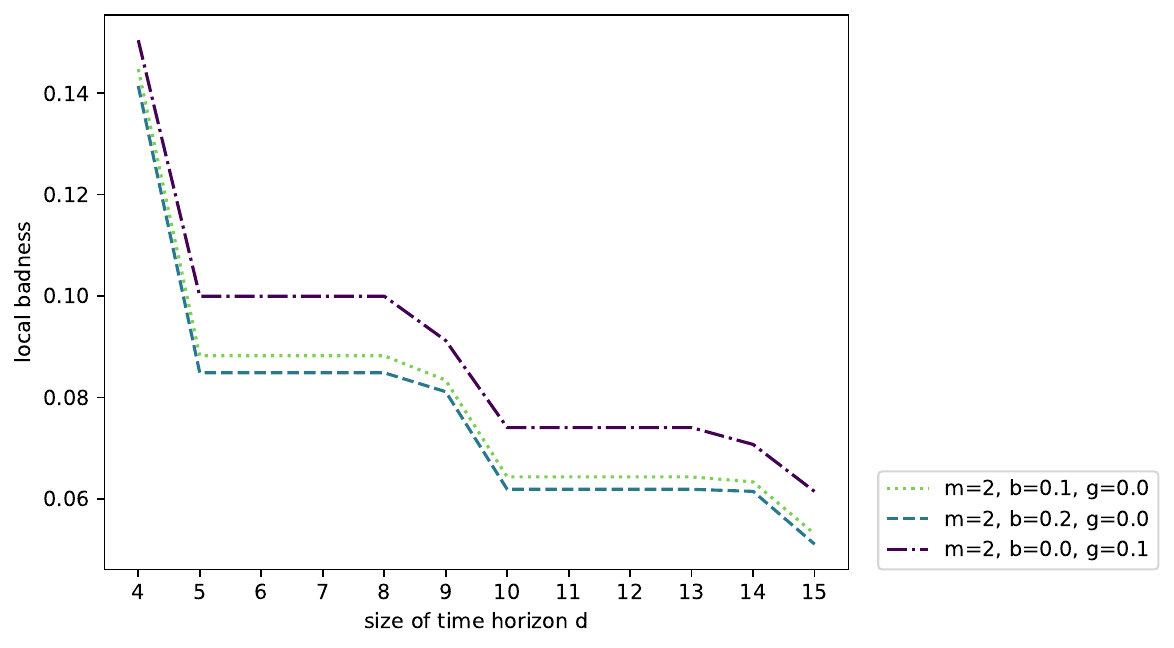}
    \includegraphics*[width=.5\textwidth]{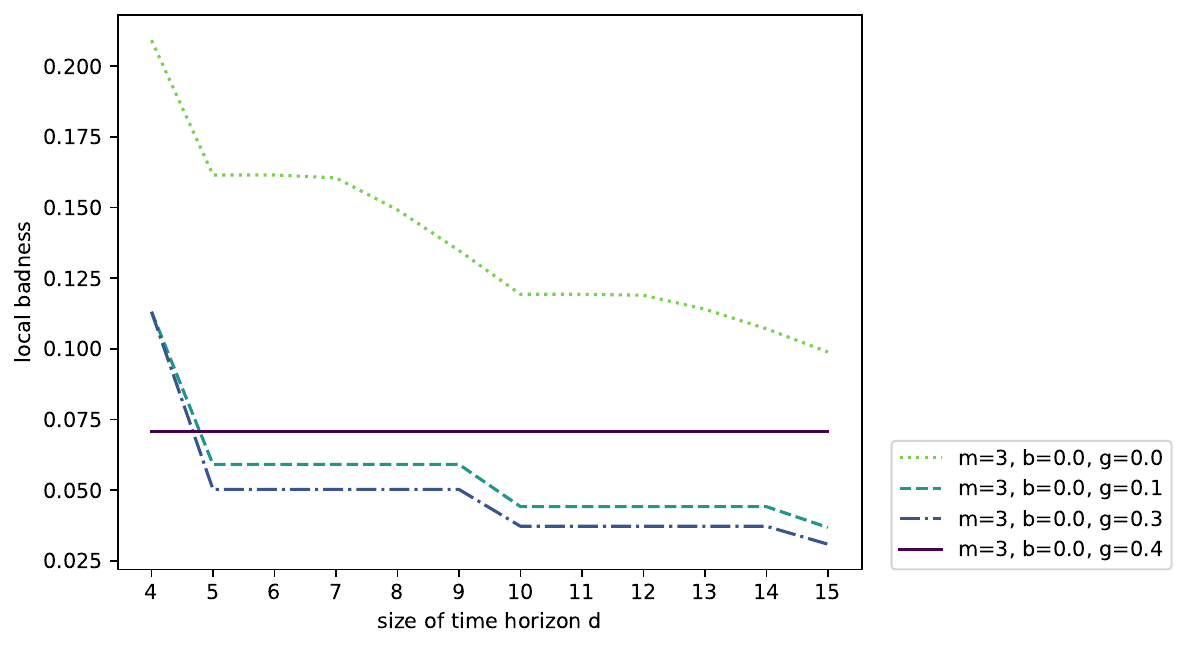}
    \caption{Strategies constructed for the graph of Fig.~1(a) in the main body of the paper, where the number of memory states allocated to the state $R$ is fixed to $m=2$ or $m=3$.}
    \label{fig-exp-I-graphs}
\end{figure*}

In Fig.~\ref{fig-exp-I-graphs}, we present plots of $\LBad^\sigma(\Distance_\nu,d)$ achieved by strategies where the number of memory states allocated to the state $R$ is fixed to $m=2$ or $m=3$.
The plots show that the same strategies determine different values of $\LBad^\sigma(\Distance_\nu,d)$ for different $d$'s, and none of them may be best for all $d$'s. 

\begin{figure*}
    \includegraphics*[width=.5\textwidth]{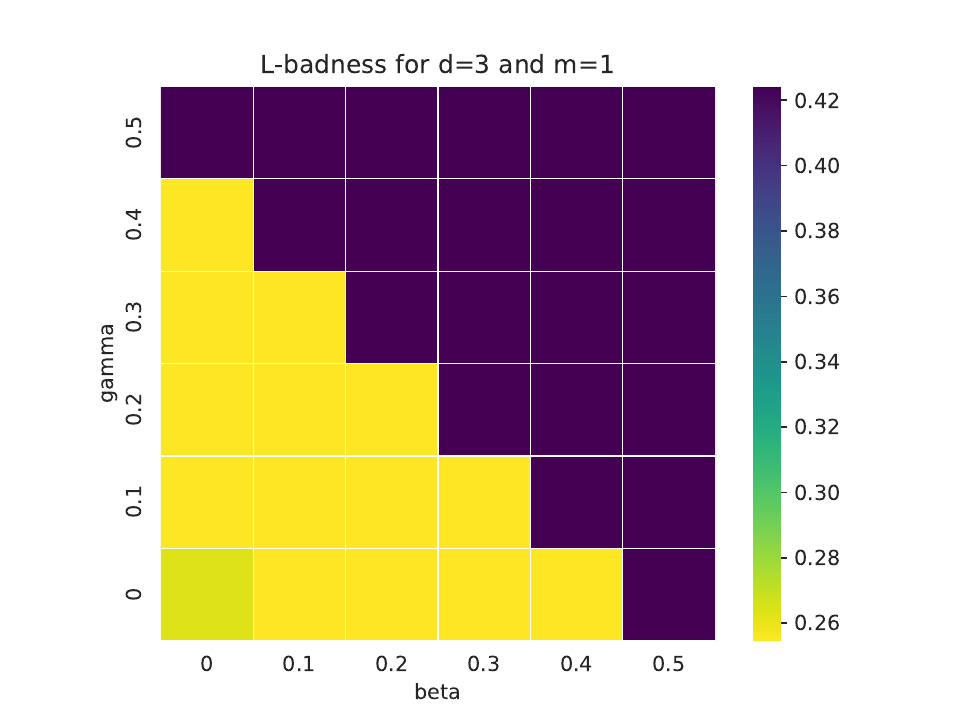}
    \includegraphics*[width=.5\textwidth]{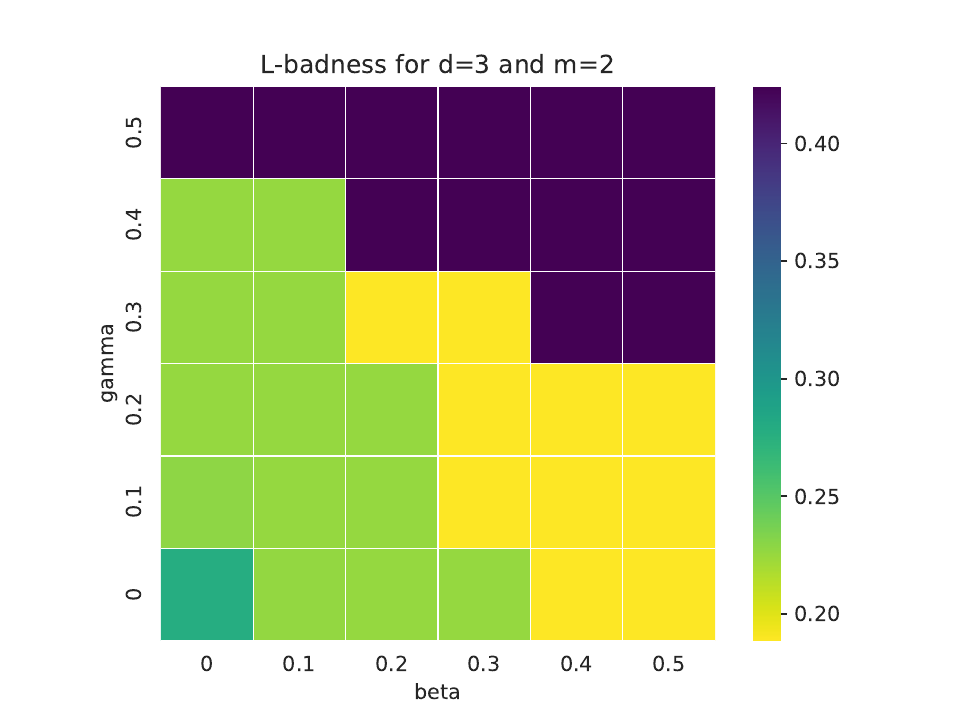}
    \includegraphics*[width=.5\textwidth]{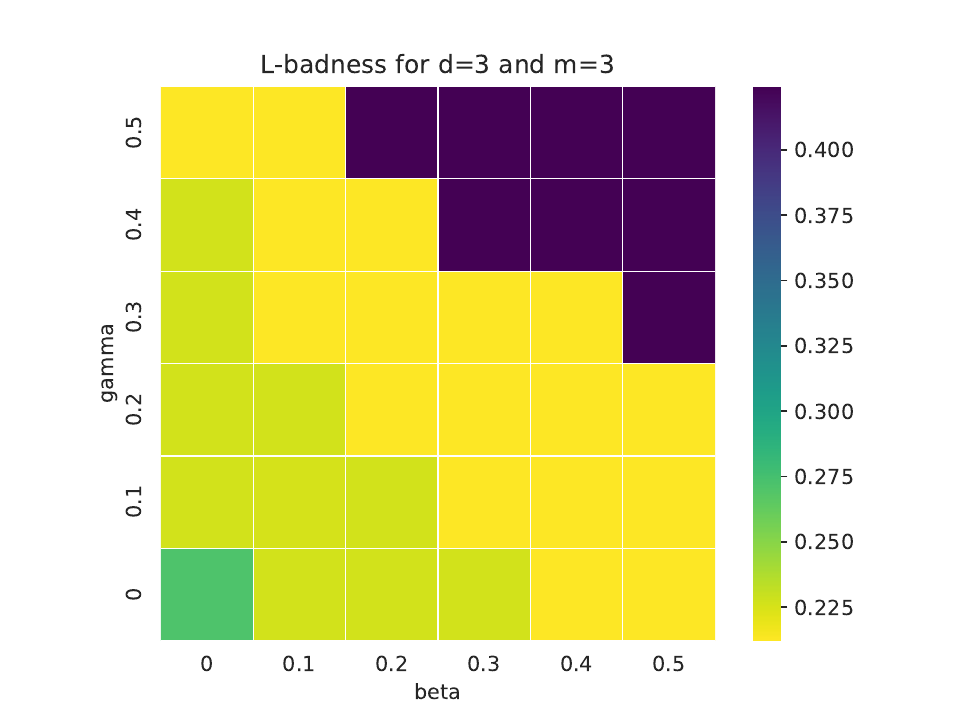}
    \includegraphics*[width=.5\textwidth]{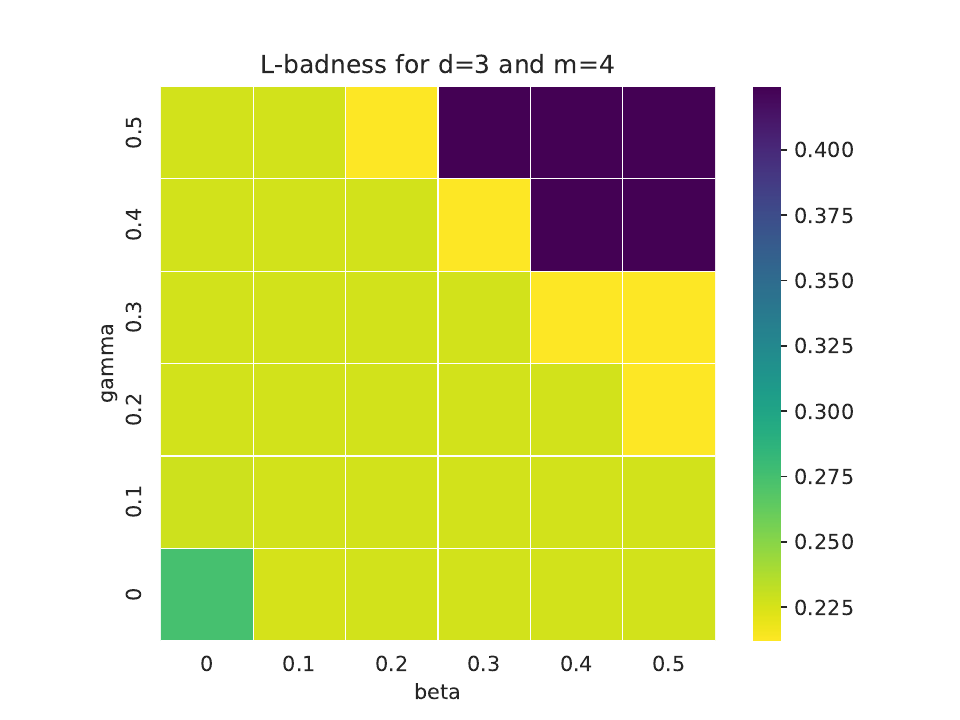}
    \caption{Strategies constructed for the graph of Fig.~1(a) in the main body of the paper, where $d =3$ and number of memory states allocated to the state $R$ ranges from $1$ to~$4$.}
    \label{fig-exp-I-graphsat3}
\end{figure*}

\begin{figure*}
    \includegraphics*[width=.5\textwidth]{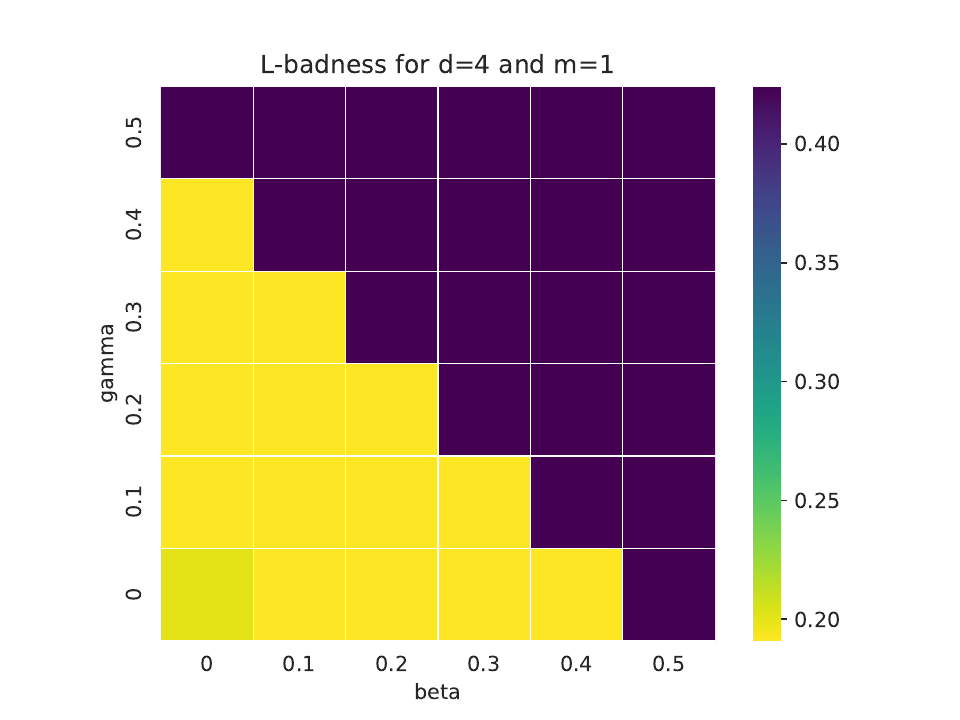}
    \includegraphics*[width=.5\textwidth]{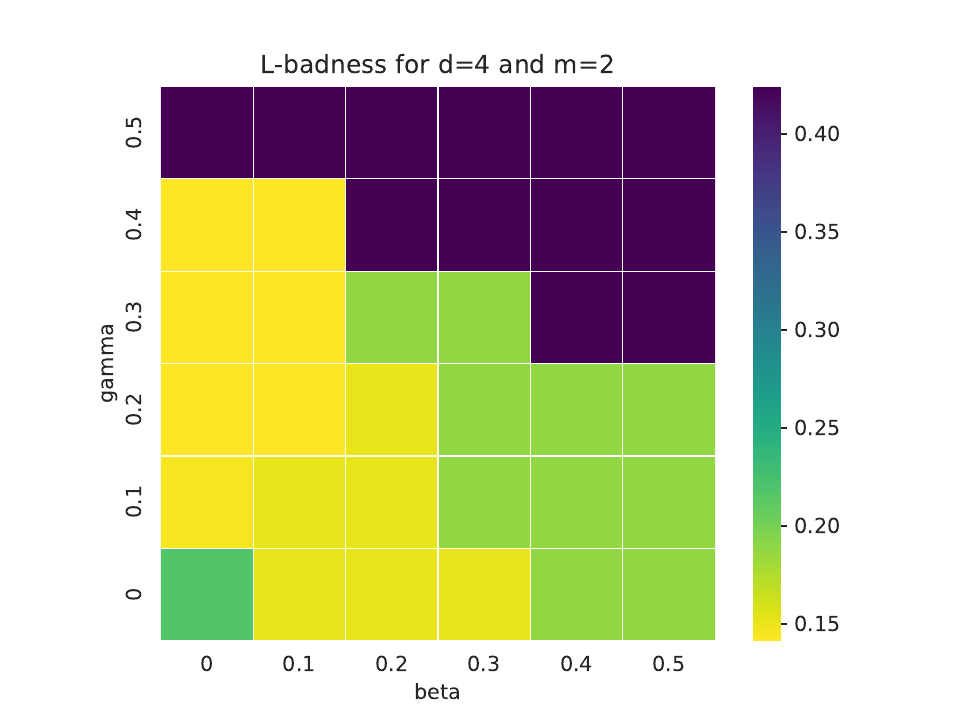}
    \includegraphics*[width=.5\textwidth]{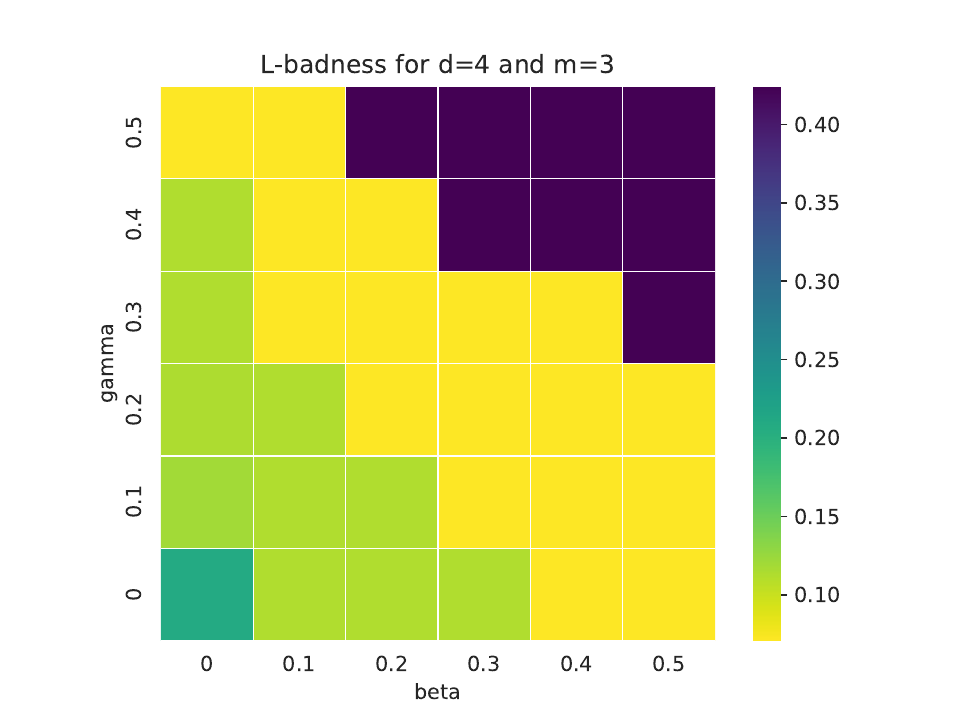}
    \includegraphics*[width=.5\textwidth]{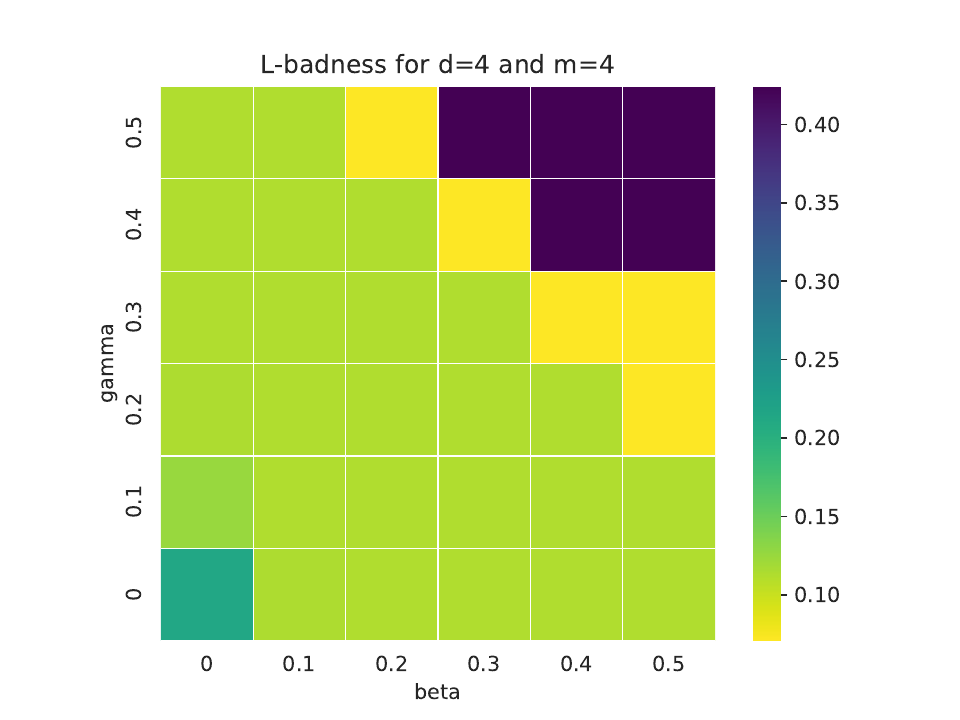}
    \caption{Strategies constructed for the graph of Fig.~1(a) in the main body of the paper, where $d =4$ and number of memory states allocated to the state $R$ ranges from $1$ to~$4$.}
    \label{fig-exp-I-graphsat4}
\end{figure*}

\begin{figure*}
    \includegraphics*[width=.5\textwidth]{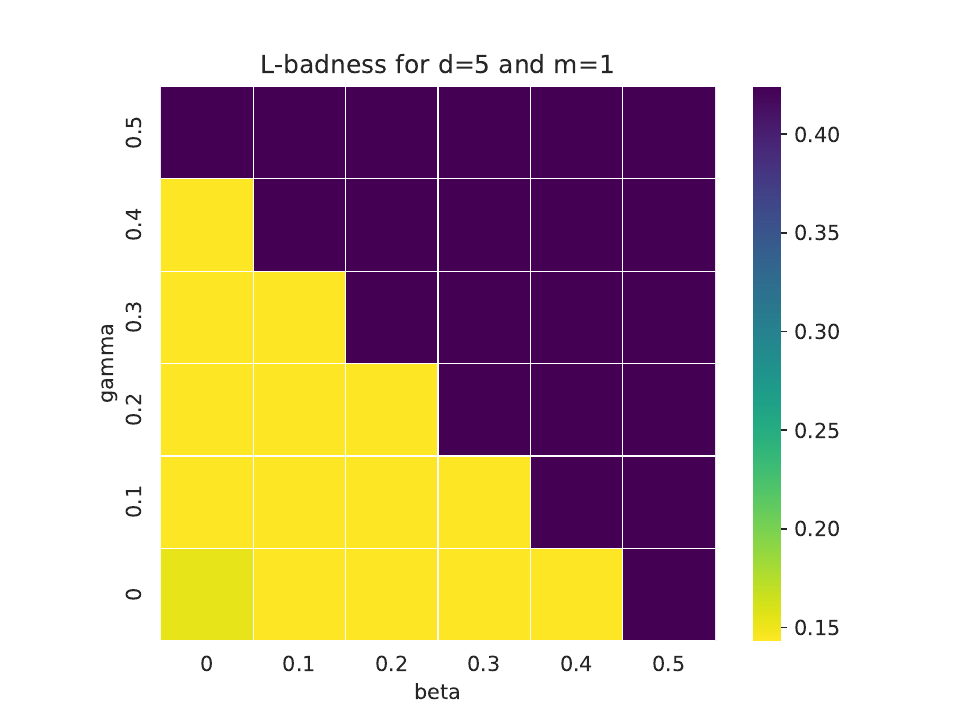}
    \includegraphics*[width=.5\textwidth]{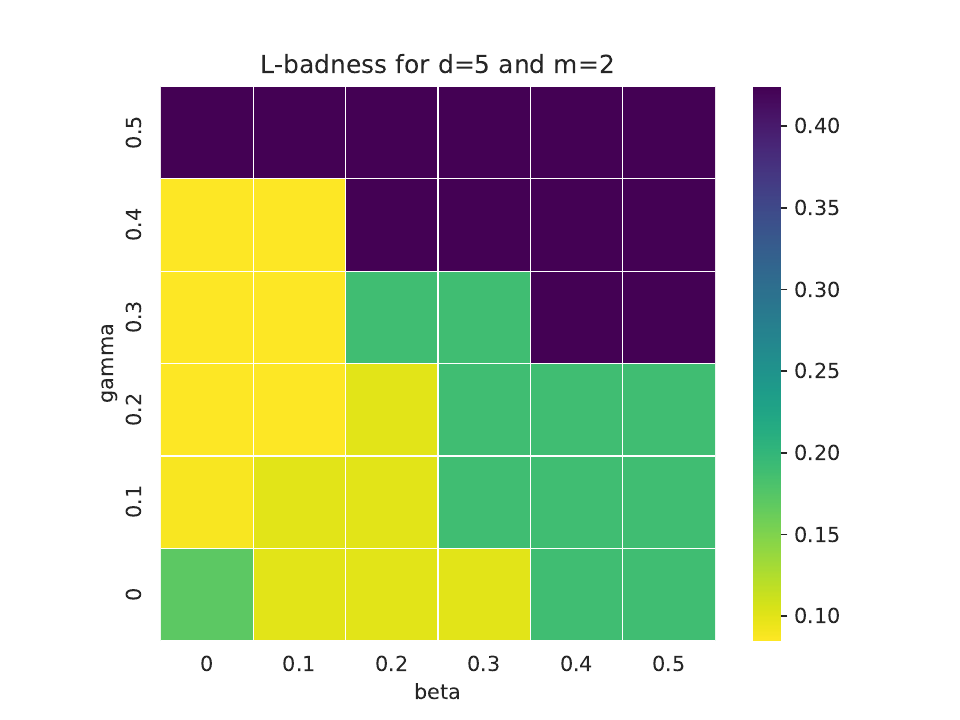}
    \includegraphics*[width=.5\textwidth]{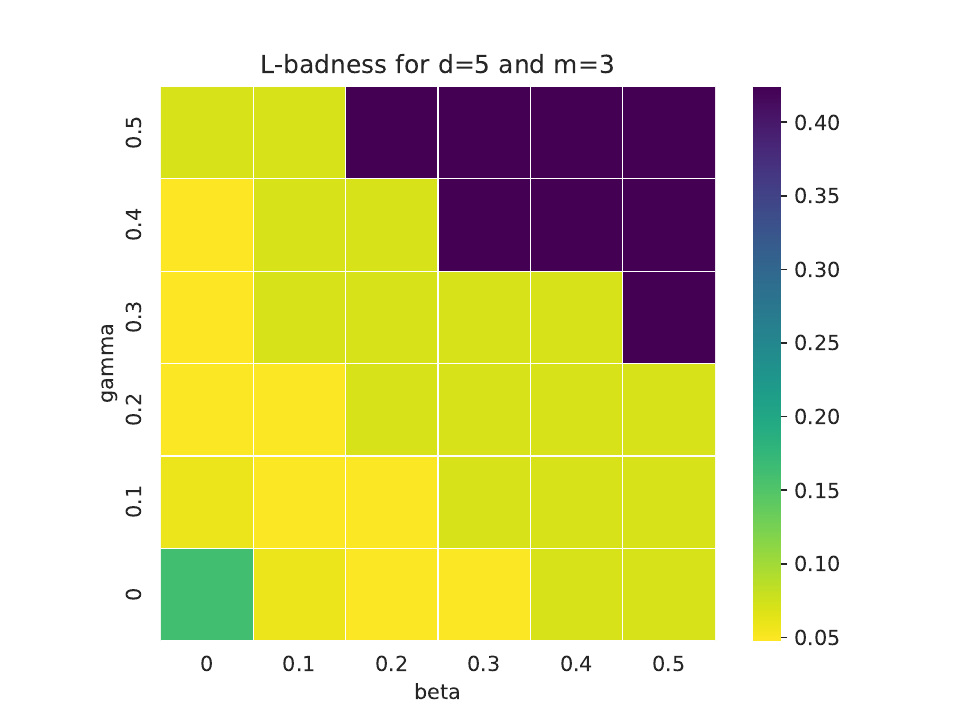}
    \includegraphics*[width=.5\textwidth]{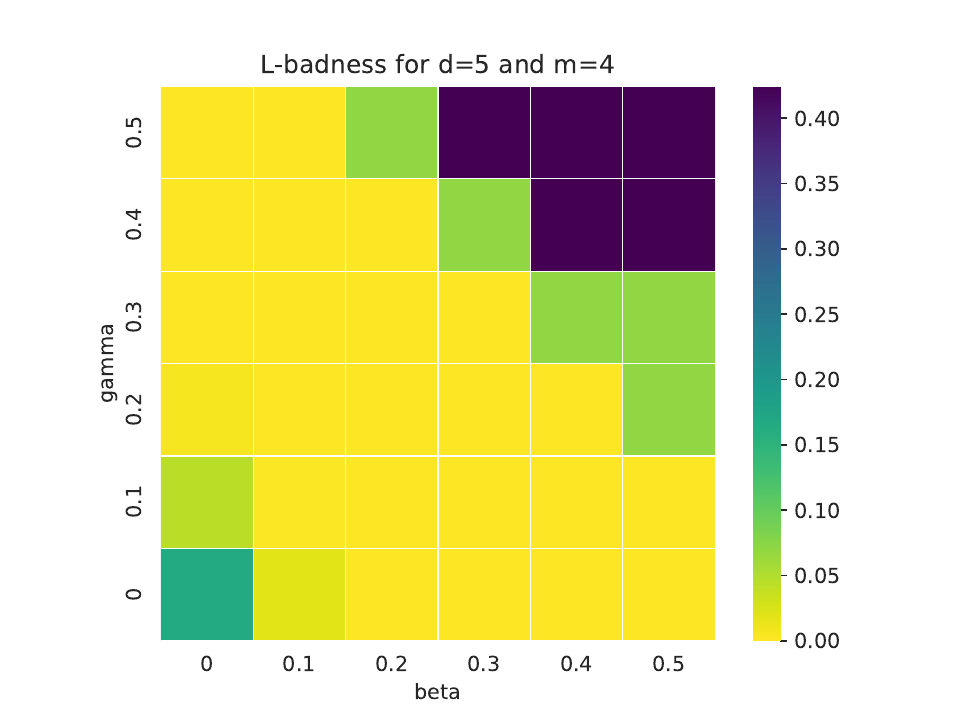}
    \caption{Strategies constructed for the graph of Fig.~1(a) in the main body of the paper, where $d =5$ and number of memory states allocated to the state $R$ ranges from $1$ to~$4$.}
    \label{fig-exp-I-graphsat5}
\end{figure*}

\begin{figure*}
    \includegraphics*[width=.5\textwidth]{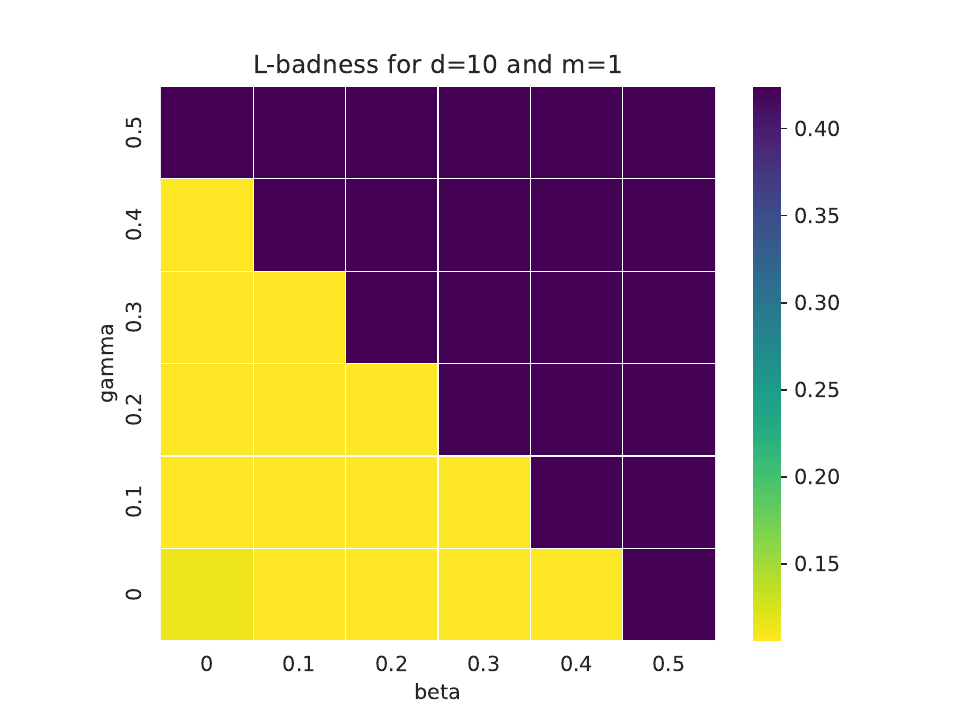}
    \includegraphics*[width=.5\textwidth]{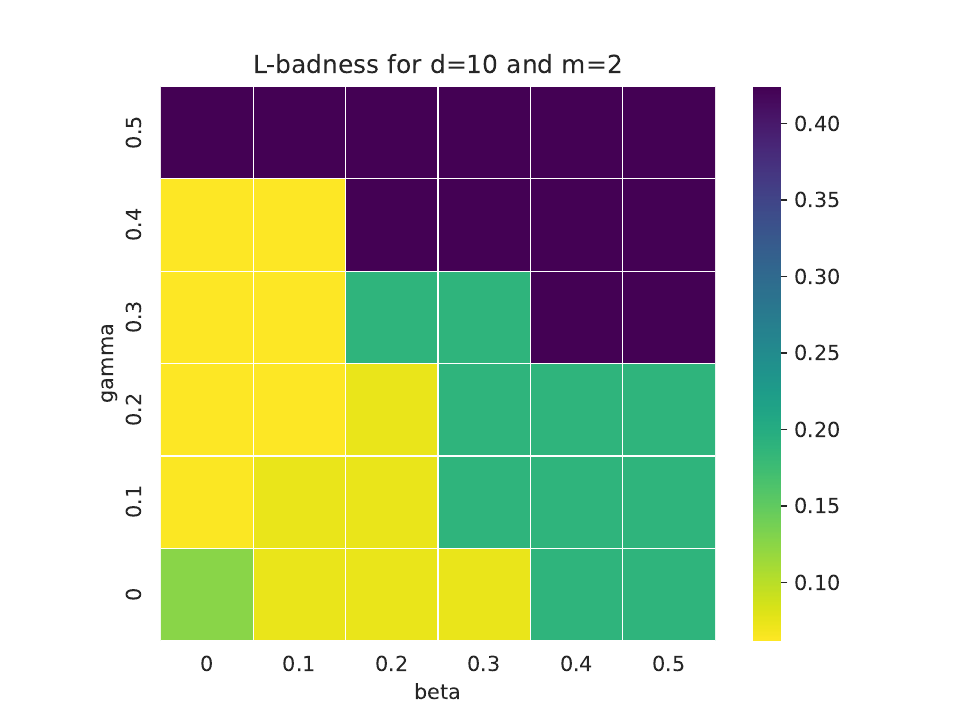}
    \includegraphics*[width=.5\textwidth]{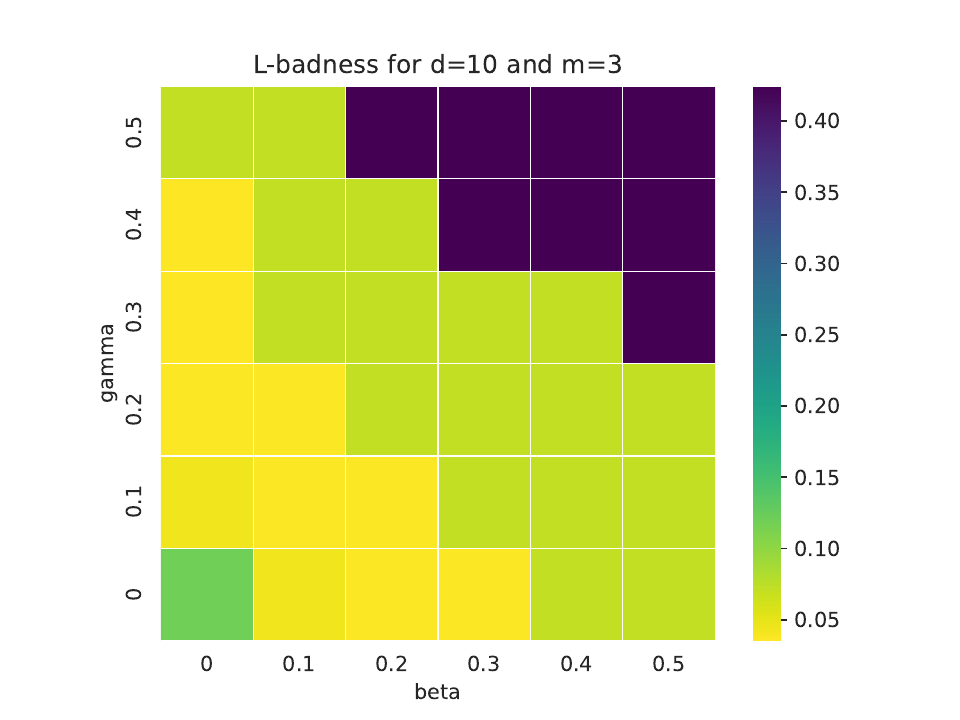}
    \includegraphics*[width=.5\textwidth]{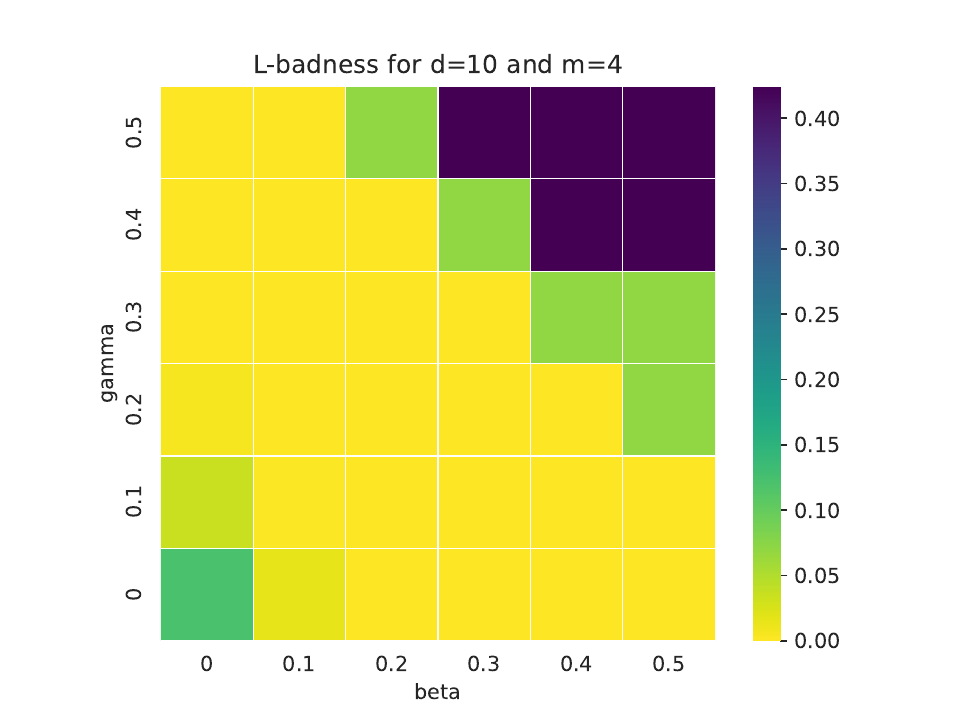}
    \caption{Strategies constructed for the graph of Fig.~1(a) in the main body of the paper, where $d =10$ and number of memory states allocated to the state $R$ ranges from $1$ to~$4$.}
    \label{fig-exp-I-graphsat10}
\end{figure*}

\begin{figure*}
    \includegraphics*[width=.5\textwidth]{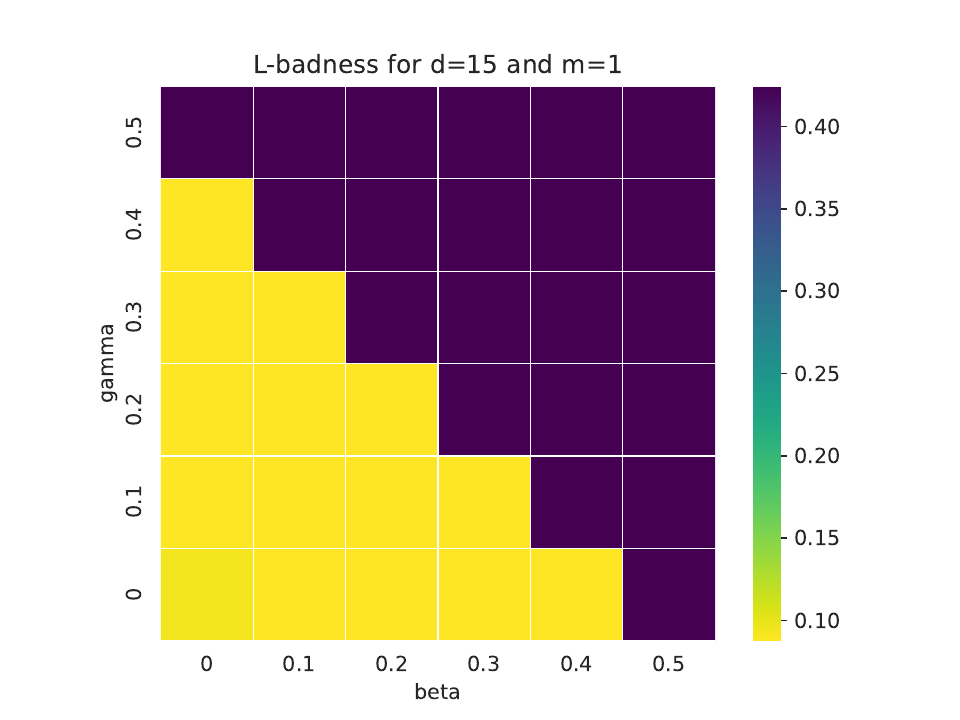}
    \includegraphics*[width=.5\textwidth]{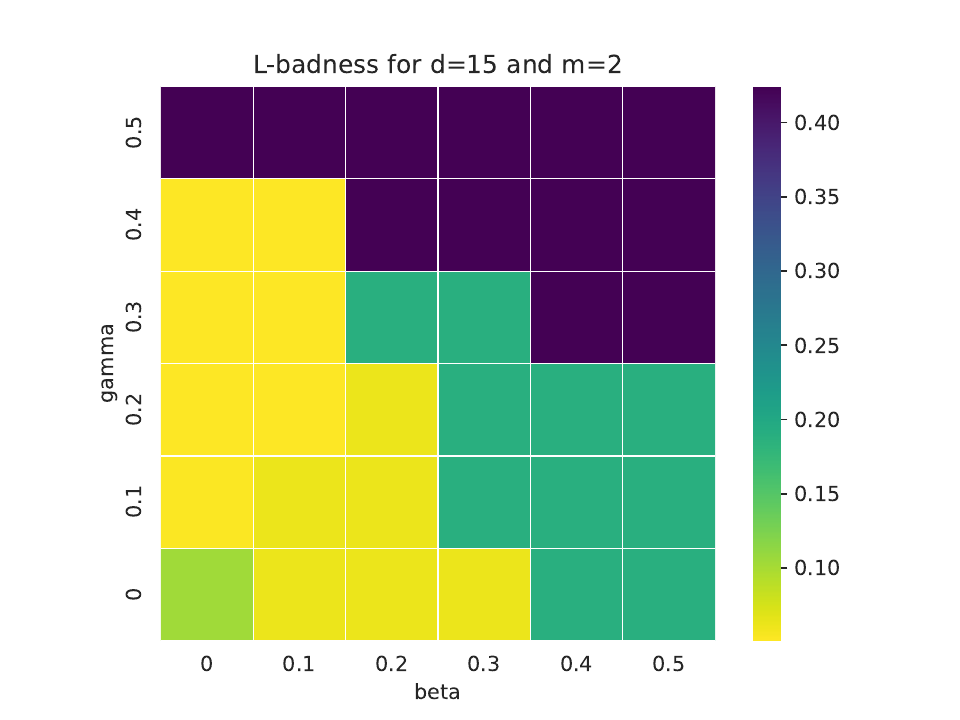}
    \includegraphics*[width=.5\textwidth]{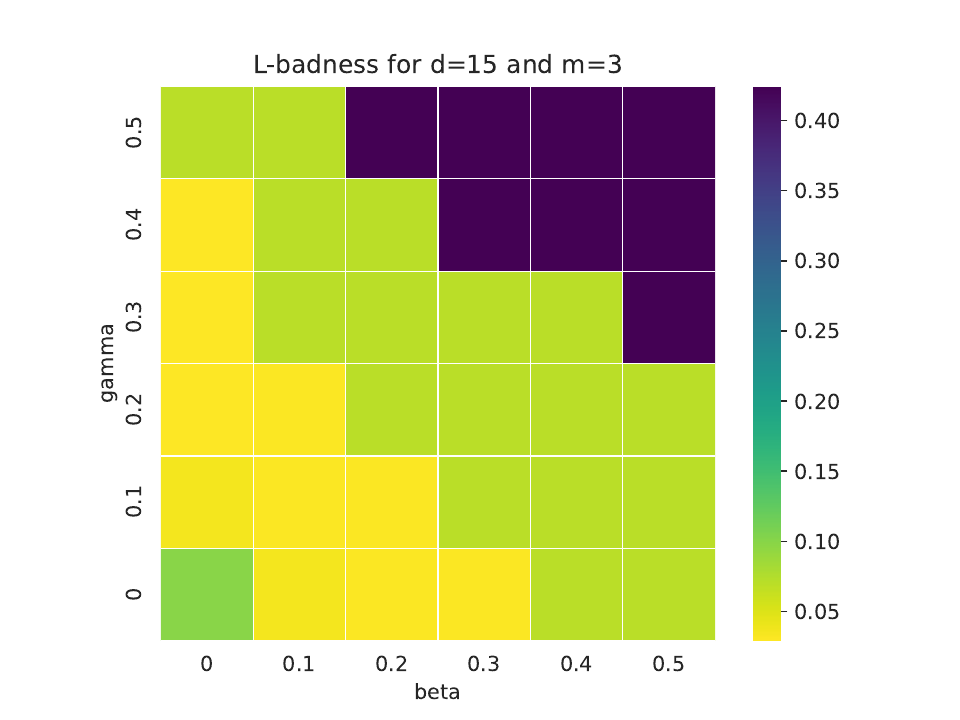}
    \includegraphics*[width=.5\textwidth]{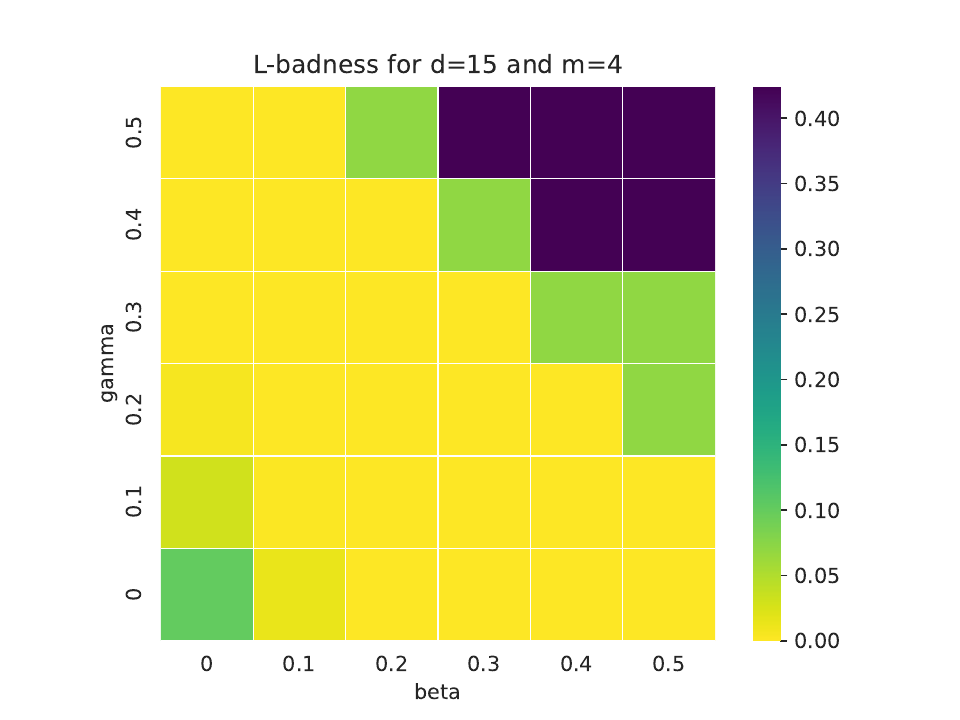}
    \caption{Strategies constructed for the graph of Fig.~1(a) in the main body of the paper, where $d =15$ and number of memory states allocated to the state $R$ ranges from $1$ to~$4$.}
    \label{fig-exp-I-graphsat15}
\end{figure*}

In Fig.~\ref{fig-exp-I-graphsat3}--\ref{fig-exp-I-graphsat15}, we analyze the quality of strategies obtained for a fixed $d \in \{3,4,5,10,15\}$ where the number of memory states allocated to the state $R$ ranges from $1$~to~$4$. For each pair of $d$ and $m$, and report $\LBad^\sigma(\Distance_\nu,d)$ for the best strategies found in $40$ runs of \LS\ with Steps set to $800$ where the parameters $\beta$ and $\gamma$ range over $\langle 0,0.5 \rangle$ with the discrete step $0.1$. All of these plots consistently show that when $\beta+\gamma$ is too large (around $0.6$), the performance of the resulting strategies is poor. Likewise, setting $\beta =\gamma =0$ (i.e., ignoring $\Penalty_1$ and $\Penalty_2$ completely) also produces strategies with poor performance. The best outcomes are generally obtained when $\beta$ and $\gamma$ are set to a certain combination of small values.

\subsection{Experiment~II}

\begin{figure*}
    \includegraphics*[width=.5\textwidth]{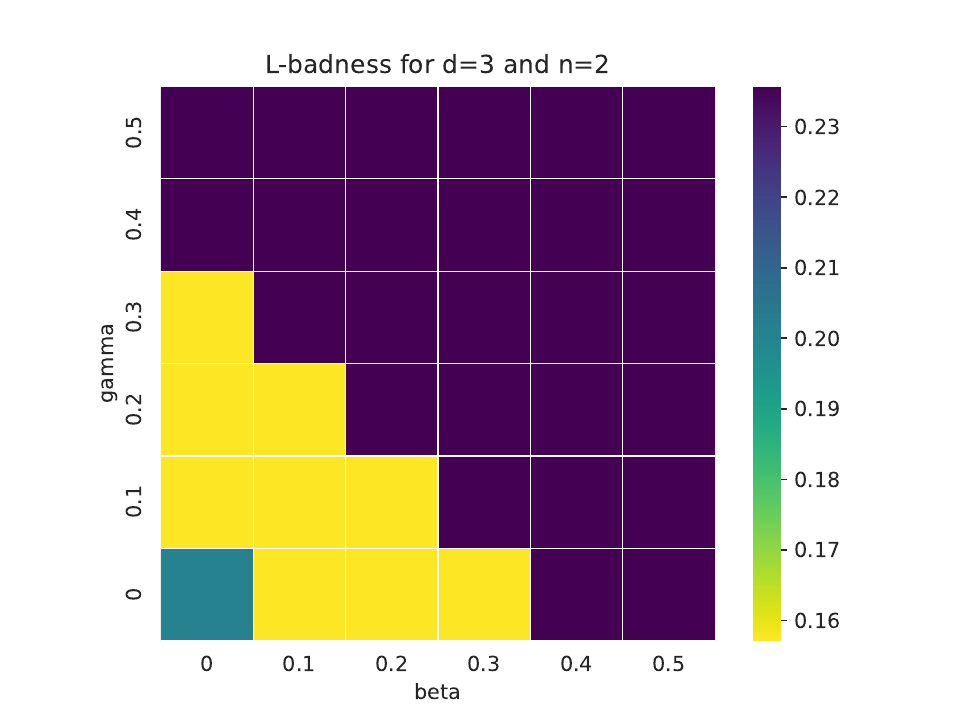}
    \includegraphics*[width=.5\textwidth]{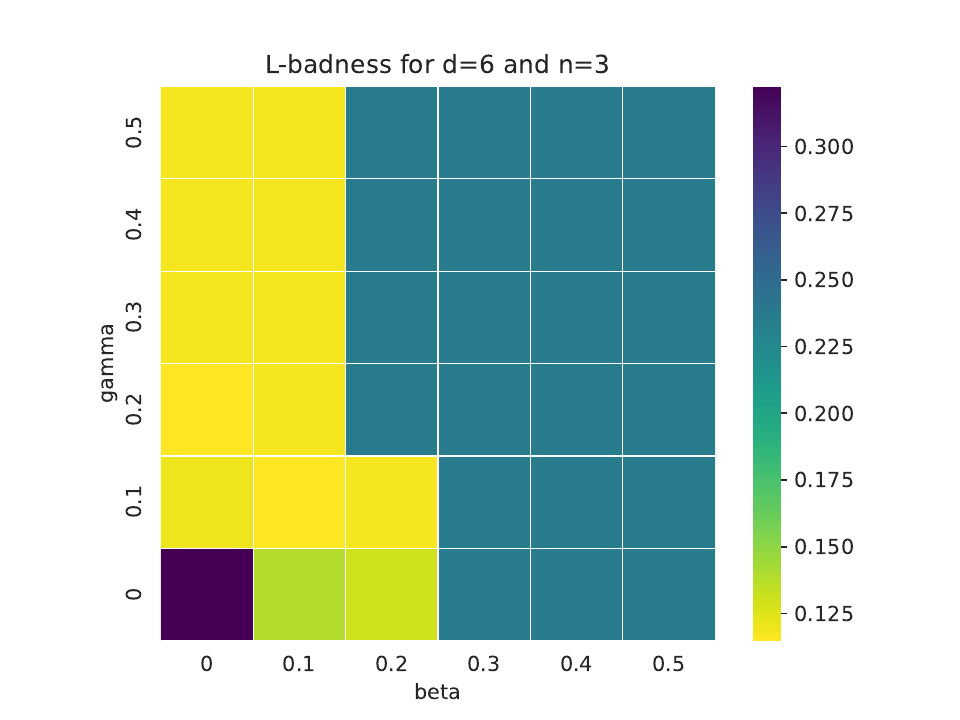}
    \includegraphics*[width=.5\textwidth]{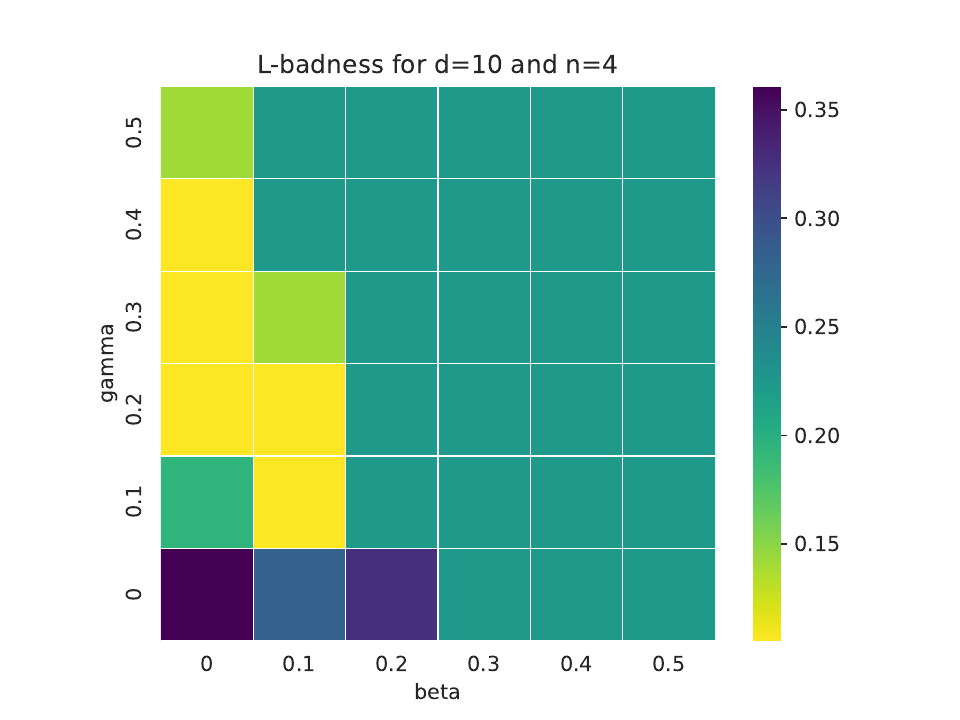}
    \includegraphics*[width=.5\textwidth]{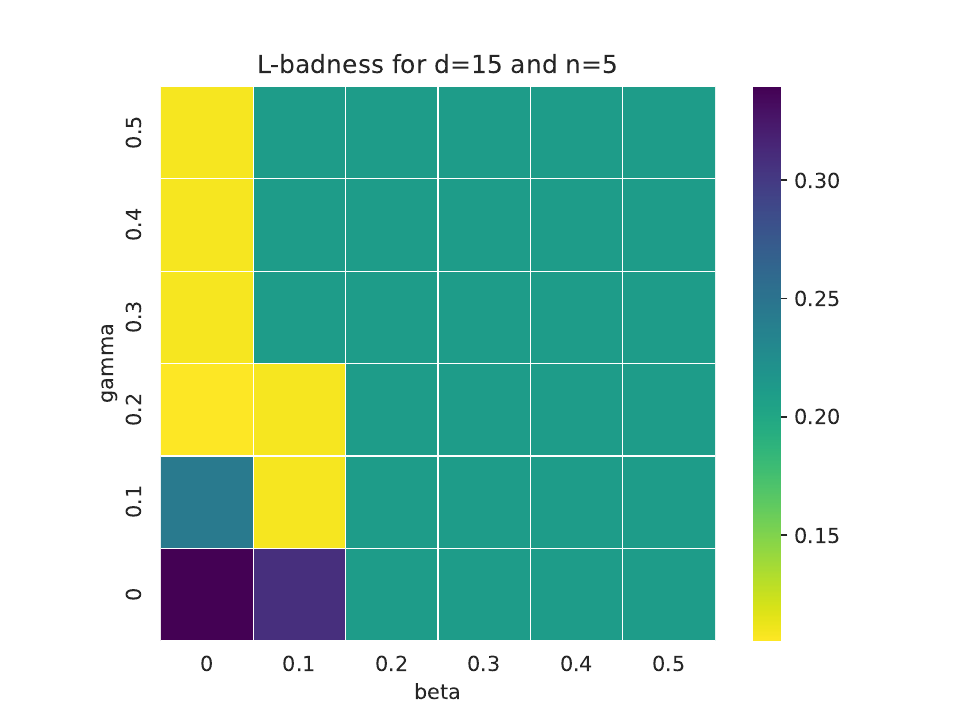}
    \includegraphics*[width=.5\textwidth]{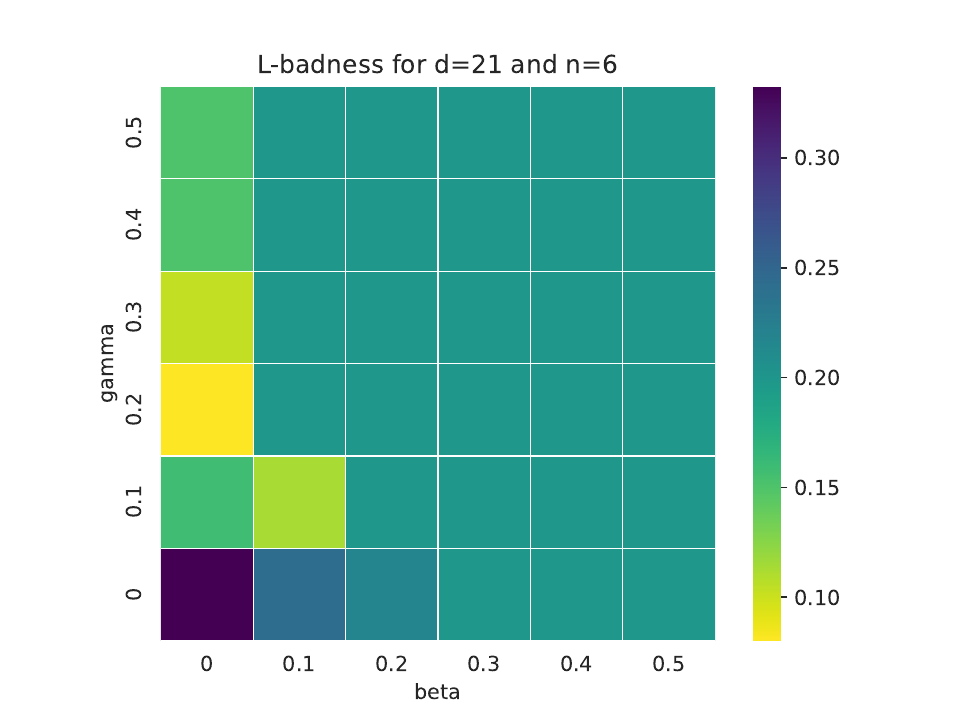}
    \caption{Strategies constructed for the graph $D_n$ in the main body of the paper, where $d = \frac{n(n+1)}{2}$.}
    \label{fig-exp-II}
\end{figure*}

In Experiment~II, we consider a set of parameterized instances $D_n$ described in Section~5.2. in the main body of the paper. In the second part of this experiment, we used \LS\ to construct the best strategies $\sigma_n$ for $D_n$ where $n \in \{2,\ldots,8\}$. This is achieved by running the \LS\ algorithm $40$ times for all $\beta,\gamma \in \langle 0,0.5 \rangle$ with the discrete step $0.1$ and collecting the strategy with the smallest value of $\Comb$. The results are summarized in Table~1 in the main body of the paper. In Fig.~\ref{fig-exp-II}, we give detailed plots of the strategy quality obtained for different combinations of the $\beta$ and $\gamma$ parameters.

Furthermore, we compared the local badness of the best strategy $\sigma_n$ computed by \LS\ against the local badness of the strategies $\pi_n$ and $\varrho_n$. The strategy $\sigma_n$ outperform $\pi_n$ and $\varrho_n$ for all $n \geq 3$, and the most significant improvement is achieved for $n=6$.
The structure of $\sigma_6$ is shown in Table~\ref{tab-sigma6}. The rows represent the probabilities of outgoing edges for every augmented vertex. Recall that $\pi_6$ and $\varrho_6$ are constructed so that the invariant distribution on the vertices $v_1,\ldots,v_6$ achieved by these strategies is equal to the ``locally desired'' distribution 
\[
\nu = (0.048, 0.095, 0.143, 0.190, 0.238, 0.286). 
\]
However, $\sigma_6$ achieves a \emph{different} invariant distribution
\[
    (0.099, 0.002, 0.100, 0.200, 0.200, 0.200)
\]
Hence, $\sigma_6$ ``trades'' the global satisfaction for the local satisfaction. Clearly, $\sigma_6$ cannot be constructed by hand.

\setlength{\tabcolsep}{1ex}
\begin{table*}[t]
    \footnotesize
    \centering
    \begin{tabular}{c|ccccccccccccccc}
    & $(v_1,1)$ & $(v_2,1)$ & $(v_2,2)$ & $(v_3,1)$ & $(v_3,2)$ & $(v_3,3)$
    & $(v_4,1)$ & $(v_4,2)$ & $(v_4,3)$ & $(v_5,1)$ & $(v_5,2)$ & $(v_5,3)$
    & $(v_6,1)$ & $(v_6,2)$ & $(v_6,3)$\\
    \cmidrule(r){1-16}
    $(v_1,1)$  & & $0.987$ & $0.013$\\
    $(v_2,1)$  & & & & & $1$\\
    $(v_2,2)$  & & $0.018$ & & $0.331$ & $0.638$ & $0.012$  \\
    $(v_3,1)$  & & & & $0.053$ & $0.172$ & $0.066$ & $0.424$ & $0.039$ & $0.245$\\
    $(v_3,2)$  & & & & & & & &  $1$ \\
    $(v_3,3)$  & & & &  $0.053$ & $0.333$ & $0.077$ & & $0.128$ & $0.408$ \\
    $(v_4,1)$  & & & & & & & & & & & & $1$ \\
    $(v_4,2)$  & & & & & &  & $1$ \\
    $(v_4,3)$  & & & & & &  & $0.021$ & $0.012$ & $0.022$ & $0.665$ & $0.279$  \\
    $(v_5,1)$  & & & & & & & & & & $0.057$ & $0.759$ & $0.096$ & $0.044$ & $0.015$ & $0.029$ \\
    $(v_5,2)$  & & & & & & & & & & & & & & & $1$ \\
    $(v_5,3)$  & & & & & & & & & & & $1$ \\
    $(v_6,1)$  & & & & & & & & & & & & & &  $1$ \\
    $(v_6,2)$  & $1$ \\ 
    $(v_6,3)$  & & & & & & & & & & & & &  $0.984$ & $0.016$ \\     
    \end{tabular}
    \caption{The structure of $\sigma_6$.}
    \label{tab-sigma6}
\end{table*}

